\documentclass[11pt]{article}

\usepackage{amsmath, amssymb, amsthm}
\usepackage{geometry}
\usepackage{setspace}
\usepackage{bm}
\usepackage{booktabs}
\usepackage{array}
\usepackage{graphicx}
\usepackage{pgf}
\usepackage{caption}
\usepackage{subcaption}
\usepackage[authoryear,round]{natbib}

\theoremstyle{definition}
\newtheorem{remark}{Remark}

\title{Dynamic Mini--Max Design and Sequential HB Inference\\
for Repeated Surveys}
\author{Siu-Ming Tam\\
Tam Data Advisory, Australia\\
\texttt{stattam@gmail.com}}
\date{}

\begin{document}

\maketitle

\begin{abstract}
This paper develops a Dynamic Mini-Max (DMM) framework for repeated
surveys comprising a Dynamic Mini-Max Design and a Sequential
Hierarchical Bayes Update (SHBU).  The DMM jointly optimises sample
size and wave overlap subject to simultaneous precision constraints for
levels and movements, a respondent burden limit and a fieldwork budget.

The methods are illustrated using 2021 Australian Census data ($t=1$)
and simulated waves $t=2,3,4$.  Both the DMM and the classical design
start from the same 5\% proportional allocation of $n^A = 42{,}018$
units.  The DMM reduces this to $n^* = 40{,}251$ while meeting all
precision constraints, achieving a cost saving of approximately 6.3\%.
Level coverage is comparable between the two designs
(maximum absolute relative error (MARE) ratio 0.844--1.263). Movement coverage diverges markedly: the DMM achieves
100\% across all 27 domain-variable cells while the classical design
achieves only 82\%--96\% (87.5\%--95.0\% nationally). The classical
confidence interval understates movement uncertainty because it
addresses sampling variance only and does not account for the model
variance component $\hat{V}_{\mathrm{mod}}$. Additional benefits of
the DMM framework --- coherent joint inference for levels and
movements, sequential updating without ad hoc composite estimator
chaining, and small area estimation --- are outlined in the paper.
\end{abstract}

\section{Introduction}
It is customary for national statistical offices (NSOs) to produce reliable estimates not only of levels at each wave, but also of movements over time for key indicators such as employment, unemployment, labour force participation, retail sales and other economic and social measures in repeated surveys. While cross-sectional surveys primarily focus on estimating the level of a characteristic at a single point in time, repeated surveys additionally require the accurate estimation of movement between successive waves. This introduces an additional design challenge because the precision of movement estimates depends not only on the sample size at each wave, but also on the degree of sample overlap between waves.

\citet{tam2026a} showed that combining multivariate Bethel \citep{bethel1989} allocation with Hierarchical Bayes (HB) small area models can substantially reduce survey sample sizes while maintaining domain-level precision and near-nominal interval coverage. That framework, the Mini-Max framework, addressed a single cross-sectional survey and focused on determining the minimum sample size required to satisfy precision requirements simultaneously across multiple variables and domains. In practice, however, many major surveys conducted by NSOs operate continuously through monthly, quarterly or annual cycles. Such surveys require a framework that extends beyond static sample allocation and incorporates the temporal evolution of the population.

Repeated surveys have been studied extensively in both the design-based and model-based literature. The studies cited here represent only a small subset of a much broader body of work, including methods for estimating movement from overlapping samples and frameworks that exploit temporal dependence across waves. Within the classical survey sampling literature, \citet{patterson1950} developed foundational theory for sampling on successive occasions with partial replacement and established covariance relationships between estimates across waves. \citet{raograham1964} extended these ideas to multi-occasion rotation designs, while \citet{bailar1975} documented the practical implications of rotation group bias in panel surveys. \citet{cochran1977} provides the standard textbook treatment of successive sampling and overlap.

More recent developments increasingly exploit model-based frameworks for repeated surveys and small area estimation. \citet{tam1987} proposed a dynamic linear model framework for repeated surveys using state-space methods and maximum likelihood estimation. That framework focused on retrospective estimation of level trajectories from completed survey data; it did not address prospective design of sample sizes and overlaps, nor did it incorporate small area estimation or movement SE constraints. State-space approaches for modelling temporal dependence are discussed extensively by \citet{harvey1989}. \citet{pfeffermannburck1990} and \citet{raoyu1994} developed approaches combining cross-sectional and time-series information for small area estimation, while \citet{pfeffermann2003} reviewed broader developments and emerging directions in model-based small area estimation. \citet{raomolina2015} provide a comprehensive treatment of model-based and time-series approaches for small area estimation.

The present paper builds on this foundation to address a complementary but
distinct problem: prospective design rather than retrospective estimation.
Given the information from wave $t-1$, what sample size $n_{h,t}$ and overlap
$m_{h,t}$ should be chosen for wave $t$ to minimise fieldwork cost while meeting
precision requirements for both levels and movements across all domains and
variables, subject to respondent burden and budget constraints?

This problem is distinct from the existing literature in the following sense.
\citet{patterson1950} and \citet{raograham1964} provide optimal overlap
formulae for classical rotation designs, but treat the total sample size as
given and do not incorporate model-based precision. Composite estimation
methods \citep{cochran1977} exploit wave-to-wave overlap for estimation but
do not address prospective design. \citet{raoyu1994} and
\citet{pfeffermannburck1990} use state-space and time-series models for
small area estimation in repeated surveys, but do not jointly optimise the
design variables $(n_{h,t}, m_{h,t})$. The DMM framework addresses all three
limitations simultaneously: it uses a Bayesian state-space model for
estimation, treats $(n_{h,t}, m_{h,t})$ as joint decision variables, and
produces coherent uncertainty quantification for both levels and movements.

To address this problem, we extend the Mini-Max framework from a static cross-sectional setting to a repeated-survey setting through a dynamic HB state-space framework comprising a sampling model, a linking model and an AR(1) propagation model. The propagation model acts as the prior distribution for the subsequent wave, allowing information from earlier waves to be carried forward through a Sequential Hierarchical Bayes Update (SHBU). Rather than treating survey waves independently, posterior uncertainty is propagated sequentially across waves.

Several extensions are required to move from the cross-sectional framework to the repeated-survey setting. First, the static framework is replaced by a sequential wave-by-wave updating structure that reflects the operational reality of ongoing surveys where future observations are unavailable at the time of design. Second, the DMM framework is formulated at the stratum level while supporting domain-level publication requirements and retaining consistency with the multivariate allocation strategy described in \citet{tam2026a}. Finally, sample overlap $m_{h,t}$ is treated as an explicit design variable rather than a fixed operational parameter.

The key contribution of this paper is the development of a DMM framework for repeated surveys that enables coherent Bayesian inference and joint optimisation of sample size and overlap. The DMM framework comprises two integrated components: a DMM Design that determines the optimal sample sizes and overlap fractions at each wave, and a SHBU that propagates posterior uncertainty across waves. The
original Mini-Max framework of \citet{tam2026a} identified the smallest
sample satisfying level CV precision constraints using HB models for a
single cross-sectional wave.  The present paper extends this to repeated
surveys by adding movement SE constraints, wave overlap as an explicit
joint decision variable, and sequential wave-by-wave updating.  The
two papers address different constraint sets --- \citet{tam2026a}
imposes level CV constraints only, while the present paper additionally
imposes movement SE constraints --- so direct comparison of achieved
sample reductions is not meaningful.

The central result is that the DMM, starting from the same 5\%
proportional allocation as the classical design ($n^A = 42{,}018$),
reduces the required sample to $n^* = 40{,}251$ while simultaneously
meeting all level CV and movement SE precision constraints, at a
fieldwork cost saving of approximately 6.3\%.  The comparison between the two
designs is therefore not primarily about which meets the level
precision constraints --- both do, at comparable sample sizes --- but
about what the DMM provides beyond constraint satisfaction: a smaller
sample through joint optimisation of $(n_{h,t}, m_{h,t})$; coherent
joint uncertainty for levels and movements; and sequential updating
without ad hoc composite estimator chaining.  It should be noted
that the DMM and the classical design address different inferential
targets for movement: the DMM movement CBI quantifies total
uncertainty including both sampling and model components, while the
classical CI targets design-based sampling variance only.  The
comparison is therefore not one of the same estimator under two
designs, but of two different inferential frameworks applied to the
same data.

The layout of the paper is as follows. Section~2 introduces the setting and
notation. Section~3 develops the three-model dynamic HB state-space structure
--- sampling, linking, and AR(1) propagation --- and consolidates all prior
distributions. Section~4 derives the full conditional distributions for the
Metropolis-within-Gibbs (MwG) sampler. Section~5 describes the SHBU component of the DMM framework, establishes its validity as a draw
from the correct joint posterior, and defines the post-design level and movement
estimators. Section~6 introduces the precision requirements, the prior inputs,
and the DMM Design component of the DMM framework. Section~7 illustrates the DMM framework using 2021 Australian Census microdata for wave 1 and simulated data for waves 2 to 4. Section~8 concludes.

\section{Setting and Notation}

Let $\mathcal{U} = \{1,\ldots,N\}$ denote a finite population of $N$ units
partitioned into $H$ non-overlapping strata indexed $h = 1,\ldots,H$, with stratum
sizes $N_1,\ldots,N_H$. The population is also partitioned into $D+1$ publication
domains indexed $d = 0,1,\ldots,D$, where $d=0$ denotes the national domain and
$d=1,\ldots,D$ denote geographic sub-national domains. Each domain $d \geq 1$ is
a union of strata: $d = \bigcup_{h \in \mathcal{H}_d} h$, where $\mathcal{H}_d$
denotes the index set of strata belonging to domain $d$, and $\mathcal{H}_0 =
\{1,\ldots,H\}$.

There are $V$ target variables $v = 1,\ldots,V$. Variables may be binary (e.g.\
employment status, unemployment status) or continuous (e.g.\ hours worked). For
binary variable $v$, let $p_{h,t}^{(v)}$ denote the true stratum proportion at
wave $t$ and $\eta_{h,t}^{(v)} = \mathrm{logit}(p_{h,t}^{(v)})$ the logit. For
continuous variable $v$, let $\theta_{h,t}^{(v)}$ denote the true stratum mean.
We use $\theta_{h,t}^{(v)}$ generically, with the understanding that for binary
variables this refers to $\eta_{h,t}^{(v)}$ on the logit scale where noted.

At wave $t$, let $n_{h,t}$ denote the actual stratum sample size and $m_{h,t}$
the number of units overlapping between waves $t-1$ and $t$ in stratum $h$, so
that $m_{h,t} \leq \min(n_{h,t-1}, n_{h,t})$. The new units recruited at wave
$t$ in stratum $h$ number $n_{h,t} - m_{h,t}$. Let $c_h^{\mathrm{new}} > 0$ and
$c_h^{\mathrm{ret}} > 0$ denote the unit costs of recruiting and interviewing a new respondent
and retaining and interviewing an existing respondent in stratum $h$ respectively, with
$c_h^{\mathrm{new}} \geq c_h^{\mathrm{ret}}$.

The domain-level mean parameter at wave $t$ is obtained by aggregating stratum
means:
\begin{equation}
    \theta_{d,t}^{(v)} = \frac{\sum_{h \in \mathcal{H}_d} N_h\,
    \theta_{h,t}^{(v)}}{N_d}, \qquad N_d = \sum_{h \in \mathcal{H}_d} N_h.
    \label{eq:domain_agg}
\end{equation}
The domain mean movement is defined as:
\begin{equation}
    \Delta_{d,t}^{(v)} = \theta_{d,t}^{(v)} - \theta_{d,t-1}^{(v)}.
    \label{eq:movement}
\end{equation}
Throughout the paper, $\hat{\theta}_{h,t}^{(v),(b)}$ denotes the $b$-th MCMC
draw of the stratum-level state --- an estimate of the true stratum mean
$\theta_{h,t}^{(v)}$ --- and $\hat{\theta}_{d,t}^{(v),(b)}$ denotes the
corresponding domain-level draw obtained by aggregating stratum draws via
\eqref{eq:domain_agg}. $\hat{\theta}_{d,t}^{(v),\mathrm{HB}}$ denotes
the HB posterior mean estimate, which is the average of the $B$ domain draws. The tilde notation
$\tilde{\theta}_{h,t}^{(v),(b)}$ is reserved for the AR(1) propagated prior
mean at the stratum level, and $\tilde{\Delta}_{d,t}^{(v),(b)}$ for the
draw-specific one-step-ahead movement prediction --- a deterministic plug-in
function of the sampled parameters, not a draw from any distribution.
Its average $\tilde{\Delta}_{d,t}^{(v),\mathrm{HB}} = \frac{1}{B}\sum_{b=1}^B
\tilde{\Delta}_{d,t}^{(v),(b)}$ is distinct from the published movement
estimator $\hat{\Delta}_{d,t}^{(v),\mathrm{HB}}$, which is the average of
the paired posterior draw differences
$(\hat{\theta}_{d,t}^{(v),(b)} - \hat{\theta}_{d,t-1}^{(v),(b)})$.

\section{The Three-Model HB State-Space Structure}

The dynamic HB model integrates three components, all specified at the stratum
level. Domain estimates are derived from stratum posteriors via
\eqref{eq:domain_agg}.

\subsection{Sampling model}

\noindent\textbf{I. Binary variables.}
Under a complex survey design, the effective sample size in stratum $h$ at wave
$t$ for binary variable $v$ is:
\begin{equation}
    n_{h,t}^{\mathrm{eff},(v)} = \frac{n_{h,t}}{\mathrm{DEFF}_h^{(v)}},
    \label{eq:eff_n}
\end{equation}
where $\mathrm{DEFF}_h^{(v)}$ is the stratum-level design effect for variable $v$.
The effective sample size accounts for the loss of precision due to clustering
relative to a simple random sample of the same size. The sampling model is:
\begin{equation}
    r_{h,t}^{(v)} \mid p_{h,t}^{(v)} \;\sim\;
    \mathrm{Binomial}\!\left(n_{h,t}^{\mathrm{eff},(v)},\, p_{h,t}^{(v)}\right),
    \label{eq:sampling_binary}
\end{equation}
where $r_{h,t}^{(v)}$ is the observed count of successes in stratum $h$ at wave
$t$. Using the effective sample size \eqref{eq:eff_n} rather than the actual size
$n_{h,t}$ ensures that the likelihood correctly reflects the precision available
from the design.

\medskip
\noindent\textbf{II. Continuous variables.}
For continuous variable $v$ in stratum $h$ at wave $t$, let
$\hat{\theta}_{h,t}^{(v)}$ denote the direct stratum sample mean. The
Fay--Herriot sampling model \citep{fayherriot1979} is:
\begin{equation}
    \hat{\theta}_{h,t}^{(v)} \mid \theta_{h,t}^{(v)} \;\sim\;
    \mathcal{N}\!\left(\theta_{h,t}^{(v)},\; \psi_{h,t}^{(v)}\right),
    \label{eq:sampling_continuous}
\end{equation}
where the sampling variance
\begin{equation}
    \psi_{h,t}^{(v)} = \mathrm{DEFF}_h^{(v)} \cdot (1 - f_{h,t})\,
    \frac{S_{h}^{2(v)}}{n_{h,t}}
    \label{eq:psi}
\end{equation}
is treated as known, with $f_{h,t} = n_{h,t}/N_h$ the sampling fraction,
$S_h^{2(v)}$ the within-stratum variance (assumed stable across waves for the purpose of sample design), and
$\mathrm{DEFF}_h^{(v)}$ the stratum-level design effect. The design effect
appears in both the binary and continuous sampling models, ensuring that complex
design features are consistently accounted for in both cases.

\subsection{Linking model}

The linking model connects the true stratum parameter to stratum-level auxiliary
covariates $\bm{z}_h \in \mathbb{R}^q$, borrowing strength across strata within
each wave.

\medskip
\noindent\textbf{I. Binary variables} (logit scale):
\begin{equation}
    \eta_{h,t}^{(v)} = \bm{z}_h^\top \bm{\beta}_t^{(v)} + u_{h,t}^{(v)},
    \qquad u_{h,t}^{(v)} \overset{\mathrm{iid}}{\sim}
    \mathcal{N}(0, \sigma_u^{2(v)}).
    \label{eq:linking_binary}
\end{equation}

\noindent\textbf{II. Continuous variables} (natural scale):
\begin{equation}
    \theta_{h,t}^{(v)} = \bm{z}_h^\top \bm{\beta}_t^{(v)} + u_{h,t}^{(v)},
    \qquad u_{h,t}^{(v)} \overset{\mathrm{iid}}{\sim}
    \mathcal{N}(0, \sigma_u^{2(v)}).
    \label{eq:linking_continuous}
\end{equation}
In both cases $\bm{\beta}_t^{(v)}$ is a vector of regression coefficients
at wave $t$ and $u_{h,t}^{(v)}$ is a stratum random effect. The variance
$\sigma_u^{2(v)} > 0$ is assumed to be common across all strata and domains.

\subsection{Propagation model}

The propagation model describes the temporal dynamics of the latent state at the stratum level 
and enters the HB model as the prior on $\theta_{h,t}^{(v)}$ given the
state at wave $t-1$. This is the key extension relative to \citet{tam2026a},
which uses a diffuse cross-sectional prior. Here the prior is informative and
time-varying, reflecting knowledge of the state at the previous wave.

\medskip
\noindent\textbf{I. Binary variables} (logit scale):
\begin{equation}
    \eta_{h,t}^{(v)} \mid \eta_{h,t-1}^{(v)},\, \phi^{(v)},\,
    \sigma_\eta^{2(v)},\, \mu^{(v)} \;\sim\;
    \mathcal{N}\!\left(\mu^{(v)} + \phi^{(v)}\!\left(\eta_{h,t-1}^{(v)}
    - \mu^{(v)}\right),\; \sigma_\eta^{2(v)}\right).
    \label{eq:prop_binary}
\end{equation}

\noindent\textbf{II. Continuous variables} (natural scale):
\begin{equation}
    \theta_{h,t}^{(v)} \mid \theta_{h,t-1}^{(v)},\, \phi^{(v)},\,
    \sigma_\eta^{2(v)},\, \mu^{(v)} \;\sim\;
    \mathcal{N}\!\left(\mu^{(v)} + \phi^{(v)}\!\left(\theta_{h,t-1}^{(v)}
    - \mu^{(v)}\right),\; \sigma_\eta^{2(v)}\right),
    \label{eq:prop_continuous}
\end{equation}
where $\mu^{(v)} \in \mathbb{R}$ is the long-run mean, $\phi^{(v)} \in (-1,1)$
is the AR(1) coefficient governing mean reversion, and $\sigma_\eta^{2(v)} > 0$
is the innovation variance. Both $\phi^{(v)}$ and $\sigma_\eta^{2(v)}$ are common
across all strata $h$ and domains $d$, borrowing strength in the time dimension
in the same way that $\sigma_u^{2(v)}$ borrows strength across strata. 

\begin{remark}
As is well known, the condition $|\phi^{(v)}| < 1$ is necessary and sufficient
for weak stationarity of the AR(1) process
\eqref{eq:prop_binary}--\eqref{eq:prop_continuous}. Under this condition the
process mean-reverts toward $\mu^{(v)}$ and the marginal variance
$\sigma_\eta^{2(v)}/(1-\phi^{2(v)})$ is finite. The boundary case
$\phi^{(v)} = 1$ yields a random walk with variance growing without bound;
$|\phi^{(v)}| > 1$ gives an explosive process. Both are empirically implausible
for socio-economic indicators such as unemployment rates, which do not drift
without bound. The \textup{Uniform}$(-1,1)$ prior on $\phi^{(v)}$ enforces
stationarity by construction.
\end{remark}

\subsection{State-space representation}

The sampling, linking and propagation models constitute a linear Gaussian
state-space model at the stratum level for continuous variables. For binary
variables, the Gaussian sampling model \eqref{eq:ss_obs} is replaced by the
Binomial likelihood \eqref{eq:sampling_binary} on the logit scale, rendering
the system nonlinear and non-Gaussian. Written generically (using
continuous-variable notation, with the understanding that the state is
$\bm{\eta}_t^{(v)}$ on the logit scale for binary variables):
\begin{align}
    \hat{\bm{\theta}}_t^{(v)} \mid \bm{\theta}_t^{(v)}
    &\;\sim\; \mathcal{N}\!\left(\bm{\theta}_t^{(v)},\; \bm{\Psi}_t^{(v)}\right)
    & &\text{(Sampling model)} \label{eq:ss_obs}\\[4pt]
    \bm{\theta}_t^{(v)}
    &= \bm{Z}\bm{\beta}_t^{(v)} + \bm{u}_t^{(v)},
    \quad \bm{u}_t^{(v)} \sim \mathcal{N}(\bm{0},\, \sigma_u^{2(v)} \bm{I})
    & &\text{(Linking model)} \label{eq:ss_link}\\[4pt]
    \bm{\theta}_t^{(v)} \mid \bm{\theta}_{t-1}^{(v)},\,\bm{\xi}^{(v)}
    &\;\sim\; \mathcal{N}\!\left(\bm{\mu}^{(v)} + \phi^{(v)}\!
    \left(\bm{\theta}_{t-1}^{(v)} - \bm{\mu}^{(v)}\right),\;
    \sigma_\eta^{2(v)} \bm{I}\right)
    & &\text{(Propagation / prior)} \label{eq:ss_prop}
\end{align}
where $\bm{\Psi}_t^{(v)} = \mathrm{diag}(\psi_{h,t}^{(v)}, h=1,...,H)$, $\bm{Z}$ is the
$H \times q$ matrix of stratum-level auxiliary covariates, $\bm{\mu}^{(v)}$ is
the $H$-vector with all entries equal to $\mu^{(v)}$, and $\bm{\xi}^{(v)} =
(\phi^{(v)}, \sigma_\eta^{2(v)}, \mu^{(v)})$ denotes the AR(1) parameter vector.
For binary variables, \eqref{eq:ss_obs} is replaced by \eqref{eq:sampling_binary}
with effective sample sizes and the state is $\bm{\eta}_t^{(v)}$ on the logit
scale. Domain estimates at each wave are obtained from stratum posteriors via
\eqref{eq:domain_agg}. The state-space equations \eqref{eq:ss_obs}--\eqref{eq:ss_prop}
describe the likelihood and propagation structure of the model; the prior
distributions on all parameters are specified in Section~\ref{sec:priors} below.

\subsection{Prior distributions}
\label{sec:priors}

All parameters in the model are assigned prior distributions. For clarity, the
complete prior specification is consolidated here.

\medskip
\noindent\textbf{I. Stratum-level states.}
At wave $t=1$ (no previous wave), the stratum states are assigned diffuse priors:
\begin{equation}
    \theta_{h,1}^{(v)} \sim \mathcal{N}(0, \tau_\theta^2), \qquad \tau_\theta^2
    \text{ large},
    \label{eq:prior_state1}
\end{equation}
following the cross-sectional HB model of \citet{tam2026a}. For $t \geq 2$, the
AR(1) propagation \eqref{eq:prop_binary}--\eqref{eq:prop_continuous} serves as the
prior, replacing \eqref{eq:prior_state1}.

\medskip
\noindent\textbf{II. Linking model parameters.}
\begin{equation}
    \bm{\beta}_t^{(v)} \sim \mathcal{N}(\bm{0}, \tau_\beta^2 \bm{I}), \quad
    \tau_\beta^2 = 10^6\; \text{(diffuse)}, \qquad
    \sigma_u^{2(v)} \sim \mathrm{Inv}\text{-}\chi^2(\nu^{(v)}, s^{2(v)}),
    \label{eq:prior_linking}
\end{equation}
where the hyperparameters $(\nu^{(v)}, s^{2(v)})$ are calibrated via the four-gate
grid search of \citet{tam2026a}. $\bm{\beta}_t^{(v)}$ is re-estimated afresh at
each wave under the diffuse prior; $\sigma_u^{2(v)}$ governs the residual
between-stratum heterogeneity not explained by the covariates.

\medskip
\noindent\textbf{III. AR(1) propagation parameters.}
\begin{equation}
    \phi^{(v)} \sim \mathrm{Uniform}(-1,1), \qquad
    \sigma_\eta^{2(v)} \sim \mathrm{Inv}\text{-}\chi^2(\nu_\eta^{(v)},
    s_\eta^{2(v)}), \qquad
    \mu^{(v)} \sim \mathcal{N}(0,\, \tau_\mu^2),
    \label{eq:prior_ar1}
\end{equation}
where $\tau_\mu^2$ is large (diffuse). The Uniform$(-1,1)$ prior on $\phi^{(v)}$
enforces stationarity and expresses ignorance about the degree of temporal
persistence. The hyperparameters $(\nu_\eta^{(v)}, s_\eta^{2(v)})$ are calibrated
from historical direct survey estimates using a moment estimator. Let
$\hat{\theta}_{d,k}^{(v)}$ denote the direct domain estimate at wave $k$ from the
existing survey programme, $\hat{\phi}^{(v)}$ a pilot AR(1) estimate obtained by
regressing $\hat{\theta}_{d,k}^{(v)}$ on $\hat{\theta}_{d,k-1}^{(v)}$ across
historical waves, and $\hat{\rho}_h^{(v)}$ the estimated between-wave correlation
of the characteristic for matched units. The general moment estimate of the
innovation variance (Appendix 1) is:
\begin{equation}
    \hat{s}_\eta^{2(v)} = \frac{1+\hat{\phi}^{(v)}}{2}
    \left[\frac{1}{K-1}\sum_{k=2}^{K}
    \left(\hat{\theta}_{d,k}^{(v)} - \hat{\theta}_{d,k-1}^{(v)}\right)^2
    - \bar{\psi}_d^{(v),\mathrm{corr}}\right],
    \label{eq:moment_scale}
\end{equation}
averaged across domains $d$, where the corrected average sampling variance is:
\begin{equation}
    \bar{\psi}_d^{(v),\mathrm{corr}} = \bar{\psi}_d^{(v)} -
    \frac{2}{K-1}\sum_{k=2}^{K}
    \frac{m_{h,k}}{n_{h,k}\, n_{h,k-1}}\,\mathrm{DEFF}_h^{(v)}\,
    S_h^{2(v)}\,\hat{\rho}_h^{(v)},
    \label{eq:psi_corr}
\end{equation}
with $\bar{\psi}_d^{(v)} = \frac{1}{K-1}\sum_{k=2}^{K}(\psi_{d,k}^{(v)} +
\psi_{d,k-1}^{(v)})$ the average sum of adjacent-wave sampling variances. The estimate $\hat{s}_\eta^{2(v)}$ is set as the prior
scale, with $\nu_\eta^{(v)}$ set to a small value (e.g.\ $\nu_\eta^{(v)} = 3$) to
keep the prior weakly informative. The prior on $\mu^{(v)}$ is kept fully diffuse:
as the long-run structural mean of the AR(1) process, $\mu^{(v)}$ should be
determined by the accumulated data rather than anchored to any single wave's
observed national mean. The two-stage procedure for selecting all hyperparameters
using classical survey truth data is discussed in Section~\ref{sec:gibbs}.  Table 1 summarises the specification of the priors for all the parameters of the models described in (14) to (17).

\medskip

\begin{table}[h!]
\centering
\caption{Complete prior specification for all parameters.}
\label{tab:priors}
\begin{tabular}{llp{6.5cm}}
\toprule
Parameter & Prior & Notes \\
\midrule
$\theta_{h,1}^{(v)}$ & $\mathcal{N}(0, \tau_\theta^2)$, $\tau_\theta^2$ large
    & Wave $t=1$ only; replaced by AR(1) for $t \geq 2$ \\[3pt]
$\bm{\beta}_t^{(v)}$ & $\mathcal{N}(\bm{0}, 10^6 \bm{I})$
    & Diffuse; re-sampled each wave \\[3pt]
$\sigma_u^{2(v)}$ & $\mathrm{Inv}\text{-}\chi^2(\nu^{(v)}, s^{2(v)})$
    & Hyperparameters $(\nu^{(v)}, s^{2(v)})$ calibrated once
      from wave $t=1$ data via four-gate grid search of
      \citet{tam2026a}; fixed across waves $t \geq 2$ \\[3pt]
$\phi^{(v)}$ & $\mathrm{Uniform}(-1,1)$
    & Enforces stationarity \\[3pt]
$\sigma_\eta^{2(v)}$ & $\mathrm{Inv}\text{-}\chi^2(\nu_\eta^{(v)}, s_\eta^{2(v)})$
    & Scale from general moment estimator \eqref{eq:moment_scale};
      $\nu_\eta^{(v)} = 3$ \\[3pt]
$\mu^{(v)}$ & $\mathcal{N}(0, \tau_\mu^2)$, $\tau_\mu^2$ large
    & Fully diffuse long-run mean \\
\bottomrule
\end{tabular}
\end{table}

\subsection{The full parameter vector}
\label{sec:params}

At wave $t$, for each variable $v$, the target of inference is the joint posterior:
\begin{equation}
    \pi\!\left(\bm{\Theta}_t^{(v)} \mid y_{1:t}\right),
    \label{eq:joint_post}
\end{equation}
where the full parameter vector $\bm{\Theta}_t^{(v)}$ comprises three groups:

\medskip
\noindent\textbf{I. Group 1 --- Stratum-level state parameters} (the primary
inferential targets):
\begin{equation}
    \left\{\theta_{h,t}^{(v)}\right\}_{h=1}^{H} \qquad
    (H \text{ scalar parameters}).
    \label{eq:state_params}
\end{equation}
For binary variables these are the logit-scale states $\eta_{h,t}^{(v)}$; for
continuous variables they are the natural-scale means $\theta_{h,t}^{(v)}$.
Domain-level estimates are obtained by aggregation via \eqref{eq:domain_agg}.

\medskip
\noindent\textbf{II. Group 2 --- Linking model parameters}:
\begin{equation}
    \bm{\beta}_t^{(v)} \in \mathbb{R}^q, \qquad \sigma_u^{2(v)} > 0
    \qquad (q + 1 \text{ scalar parameters}).
    \label{eq:linking_params}
\end{equation}
These govern cross-sectional borrowing of strength across strata at wave $t$.

\medskip
\noindent\textbf{III. Group 3 --- AR(1) propagation parameters}:
\begin{equation}
    \bm{\xi}^{(v)} = \left(\phi^{(v)},\; \sigma_\eta^{2(v)},\; \mu^{(v)}\right)
    \qquad (3 \text{ scalar parameters}).
    \label{eq:ar1_params}
\end{equation}
These govern temporal dynamics and are common across all strata and domains.

\medskip
\noindent The complete parameter vector at wave $t$ for variable $v$ is:
\begin{equation}
    \bm{\Theta}_t^{(v)} = \left(\left\{\theta_{h,t}^{(v)}\right\}_{h=1}^{H},\;
    \bm{\beta}_t^{(v)},\; \sigma_u^{2(v)},\; \phi^{(v)},\;
    \sigma_\eta^{2(v)},\; \mu^{(v)}\right),
    \label{eq:full_param}
\end{equation}
comprising $H + q + 4$ scalar parameters per variable per wave. The dimension is
fixed across waves; it does not grow with $t$. In the simulated labour force survey example
of \citet{tam2026a} with $H = 55$ strata and $q = 2$ covariates,
this gives 61 parameters per variable per wave.

\section{Full Conditional Distributions}
\label{sec:gibbs}

The joint posterior $\pi(\bm{\Theta}_t^{(v)} \mid y_t, \{\hat{\theta}_{h,t-1}^{(v),(b)}\})$
is sampled via a MwG sampler, which iterates
through the full conditional distribution of each parameter block given all other
parameters and the data. This section derives the key full conditionals, justifying
the applicability of MwG sampling and identifying where MwG
steps are required. The conjugate Gaussian and Inv-$\chi^2$ full
conditionals follow standard results in \citet{gelmanetal2014};
the Bayesian AR(1) regression and stationarity-constrained sampling
follow \citet{hamilton1994}.

Throughout, we condition on the fixed draw $\{\hat{\theta}_{h,t-1}^{(v),(b)}\}_{h=1}^H$
from wave $t-1$, suppress the superscripts $(v)$ and $(b)$ for notational brevity,
and write $\tilde{\theta}_{h} = \mu + \phi(\theta_{h,t-1} - \mu)$ for the AR(1)
prior mean for stratum $h$.

\medskip
\noindent\textbf{I. Conditional 1: Stratum states, continuous variable.}

For continuous variable $v$, given all other parameters, the full conditional
for $\theta_{h,t}$ is obtained by multiplying three Gaussian kernels: the
sampling likelihood \eqref{eq:sampling_continuous}, the linking model density
\eqref{eq:linking_continuous}, and the AR(1) prior \eqref{eq:prop_continuous}.
Since the linking model specifies $\theta_{h,t} = \bm{z}_h^\top\bm{\beta}_t +
u_{h,t}$ with $u_{h,t} \sim \mathcal{N}(0, \sigma_u^2)$, it implies
$\theta_{h,t} \mid \bm{\beta}_t, \sigma_u^2 \sim \mathcal{N}(\bm{z}_h^\top
\bm{\beta}_t,\, \sigma_u^2)$, which is a genuine probability distribution for
$\theta_{h,t}$ contributing precision $\sigma_u^{-2}$ centred on
$\bm{z}_h^\top\bm{\beta}_t$. The full conditional is therefore proportional to:
\begin{equation}
    \pi(\theta_{h,t} \mid \cdots) \;\propto\;
    \mathcal{N}(\hat{\theta}_{h,t},\,\psi_{h,t}) \times
    \mathcal{N}(\bm{z}_h^\top\bm{\beta}_t,\,\sigma_u^2) \times
    \mathcal{N}(\tilde{\theta}_h,\,\sigma_\eta^2),
    \notag
\end{equation}
and multiplying three Gaussian kernels gives the conjugate result:
\begin{equation}
    \theta_{h,t} \mid \cdots \;\sim\; \mathcal{N}\!\left(
    \frac{\psi_{h,t}^{-1}\,\hat{\theta}_{h,t} + \sigma_u^{-2}\,\bm{z}_h^\top\bm{\beta}_t
    + \sigma_\eta^{-2}\,\tilde{\theta}_{h}}
    {\psi_{h,t}^{-1} + \sigma_u^{-2} + \sigma_\eta^{-2}},\;\;
    \frac{1}{\psi_{h,t}^{-1} + \sigma_u^{-2} + \sigma_\eta^{-2}}\right).
    \label{eq:gibbs_theta_cont}
\end{equation}
This is a precision-weighted average of three sources of information: the direct
survey estimate $\hat{\theta}_{h,t}$ (precision $\psi_{h,t}^{-1}$), the linking
model prediction $\bm{z}_h^\top\bm{\beta}_t$ (precision $\sigma_u^{-2}$), and
the AR(1) propagated prior $\tilde{\theta}_h$ (precision $\sigma_\eta^{-2}$).
When the sampling variance $\psi_{h,t}$ is small (large sample), the direct
estimate dominates. When $\sigma_u^2$ is small (tight linking model), the
covariate prediction dominates. When $\sigma_\eta^2$ is small (strong temporal
signal), the AR(1) prediction dominates. All three components are sampled
jointly in the Gibbs chain, so this conditional is used at each iteration with
the current Gibbs values of $\bm{\beta}_t$ and $\sigma_u^2$.

\medskip
\noindent\textbf{II. Conditional 2: Stratum states, binary variable.}

For binary variable $v$, the full conditional for $\eta_{h,t} =
\mathrm{logit}(p_{h,t})$ is obtained by multiplying three factors: the binomial
sampling likelihood \eqref{eq:sampling_binary}, the linking model density
\eqref{eq:linking_binary} which gives $\eta_{h,t} \mid \bm{\beta}_t, \sigma_u^2
\sim \mathcal{N}(\bm{z}_h^\top\bm{\beta}_t, \sigma_u^2)$, and the Gaussian
AR(1) prior \eqref{eq:prop_binary}. The conditional is:
\begin{equation}
    \pi(\eta_{h,t} \mid \cdots) \;\propto\;
    \mathrm{Binomial}\!\left(r_{h,t}^{\mathrm{eff}};\; n_{h,t}^{\mathrm{eff}},\;
    \mathrm{logistic}(\eta_{h,t})\right) \times
    \mathcal{N}\!\left(\eta_{h,t};\; \bm{z}_h^\top\bm{\beta}_t,\; \sigma_u^2\right)
    \times
    \mathcal{N}\!\left(\eta_{h,t};\; \tilde{\eta}_{h},\; \sigma_\eta^2\right),
    \label{eq:gibbs_eta}
\end{equation}
which is not Gaussian and has no closed form and requires
a non-exact sampling step. Two standard approaches are available:
MwG \citep{tierney1994} and
P\'{o}lya-Gamma data augmentation \citep{polsonetal2013}.
Either approach leaves the target posterior \eqref{eq:gibbs_eta} invariant.  In this paper, the MwG approach is used.

\begin{remark}
The random effect $u_{h,t}^{(v)}$ does not appear as a separate entry in the parameter vector $\bm{\Theta}_t^{(v)}$ \eqref{eq:full_param}. It is an implicit function of the sampled parameters: given the current Gibbs iterate $\bm{\beta}_t$, the residual $u_{h,t} = \theta_{h,t} - \bm{z}_h^\top\bm{\beta}_t$ is reconstructed deterministically at each iteration. Its distributional contribution — $u_{h,t} \sim \mathcal{N}(0, \sigma_u^2)$ — is fully accounted for through the linking model density factor $\mathcal{N}(\bm{z}_h^\top\bm{\beta}_t, \sigma_u^2)$ in the full conditional \eqref{eq:gibbs_theta_cont}, which contributes the precision term $\sigma_u^{-2}$ to the posterior weights.
\end{remark}

\medskip
\noindent\textbf{III. Conditional 3: Regression coefficients $\bm{\beta}_t$.}

Given all stratum states $\{\theta_{h,t}\}$, the linking model
\eqref{eq:linking_continuous} implies:
\begin{equation}
    \theta_{h,t} - u_{h,t} = \bm{z}_h^\top \bm{\beta}_t, \qquad h = 1,\ldots,H,
\end{equation}
which is a standard Bayesian linear regression of $(\theta_{h,t} - u_{h,t})$ on
$\bm{z}_h$. With the diffuse Gaussian prior $\bm{\beta}_t \sim \mathcal{N}(\bm{0},
\tau_\beta^2 \bm{I})$ and the Gaussian linking model, the full conditional is:
\begin{equation}
    \bm{\beta}_t \mid \cdots \;\sim\; \mathcal{N}\!\left(
    \left(\bm{Z}^\top \bm{Z} + \frac{\sigma_u^2}{\tau_\beta^2}\bm{I}\right)^{-1}
    \bm{Z}^\top \bm{\theta}_t^*,\;\;
    \sigma_u^2 \left(\bm{Z}^\top \bm{Z} + \frac{\sigma_u^2}{\tau_\beta^2}\bm{I}
    \right)^{-1}\right),
    \label{eq:gibbs_beta}
\end{equation}
where $\bm{\theta}_t^* = \bm{\theta}_t - \bm{u}_t$ is the vector of
covariate-adjusted stratum states, with residuals $u_{h,t} = \theta_{h,t} -
\bm{z}_h^\top\bm{\beta}_t$ reconstructed from the current Gibbs iterates of
$\bm{\theta}_t$ and $\bm{\beta}_t$. As $\tau_\beta^2 \to \infty$, this reduces
to the GLS estimator. This conditional is Gaussian and sampled exactly.

\medskip
\noindent\textbf{IV. Conditional 4: Linking model variance $\sigma_u^2$.}

Given all stratum states $\{\theta_{h,t}\}$ and $\bm{\beta}_t$, the residuals
$u_{h,t} = \theta_{h,t} - \bm{z}_h^\top \bm{\beta}_t$ are observed. The
conjugate prior $\sigma_u^2 \sim \mathrm{Inv}\text{-}\chi^2(\nu, s^2)$ and the
Gaussian linking model give:
\begin{equation}
    \sigma_u^2 \mid \cdots \;\sim\; \mathrm{Inv}\text{-}\chi^2\!\left(
    \nu + H,\;\; \frac{\nu s^2 + \sum_{h=1}^H u_{h,t}^2}{\nu + H}\right).
    \label{eq:gibbs_sigma_u}
\end{equation}
This is sampled exactly.

\medskip
\noindent\textbf{V. Conditional 5: AR(1) parameters $(\phi, \sigma_\eta^2, \mu)$.}

Given all current stratum states
$\{\theta_{h,t}\}$ and the fixed previous-wave states $\{\theta_{h,t-1}\}$, the
AR(1) equation is:
\begin{equation}
    \theta_{h,t} = \mu + \phi(\theta_{h,t-1} - \mu) + \eta_h,
    \qquad \eta_h \overset{\mathrm{iid}}{\sim} \mathcal{N}(0, \sigma_\eta^2),
    \quad h = 1,\ldots,H.
    \label{eq:ar1_regression}
\end{equation}
This is a linear regression with $H$ observations. To obtain closed-form Gibbs
conditionals, we reparametrise:
\begin{equation}
    \alpha = \mu(1-\phi), \qquad \phi, \qquad \sigma_\eta^2,
    \label{eq:reparam}
\end{equation}
so that \eqref{eq:ar1_regression} becomes $\theta_{h,t} = \alpha +
\phi\,\theta_{h,t-1} + \eta_h$, a standard Bayesian AR(1) regression without
the nonlinear $\mu$ term. The reparametrised equation is linear in $(\alpha,
\phi)$ for fixed $\sigma_\eta^2$, giving:

\medskip
\noindent\textbf{VI. Conditional for $(\alpha, \phi)$ given $\sigma_\eta^2$:}
With a diffuse bivariate Gaussian prior on $(\alpha, \phi)$ (restricted to
$|\phi| < 1$), the full conditional is a truncated bivariate Gaussian:
\begin{equation}
    \begin{pmatrix}\alpha \\ \phi \end{pmatrix} \;\Bigg|\; \sigma_\eta^2,\cdots
    \;\sim\; \mathcal{TN}\!\left(\hat{\bm{\gamma}},\;\; \sigma_\eta^2
    (\bm{X}^\top \bm{X})^{-1};\;\; |\phi| < 1\right),
    \label{eq:gibbs_alpha_phi}
\end{equation}
where $\bm{X} = [\bm{1},\, \bm{\theta}_{t-1}]$ is the $H \times 2$ design matrix
with columns of ones and previous-wave states, $\hat{\bm{\gamma}} = (\bm{X}^\top
\bm{X})^{-1}\bm{X}^\top \bm{\theta}_t$ is the OLS estimate, and $\mathcal{TN}$
denotes a Gaussian truncated to $|\phi| < 1$. The truncation $|\phi^{(v)}| < 1$ is enforced by sampling from the
truncated Gaussian \eqref{eq:gibbs_alpha_phi} using standard
inverse-CDF methods.

\medskip
\noindent\textbf{VII. Conditional for $\sigma_\eta^2$ given $(\alpha, \phi)$:}
With conjugate prior $\sigma_\eta^2 \sim \mathrm{Inv}\text{-}\chi^2(\nu_\eta,
s_\eta^2)$:
\begin{equation}
    \sigma_\eta^2 \mid \alpha, \phi, \cdots \;\sim\;
    \mathrm{Inv}\text{-}\chi^2\!\left(\nu_\eta + H,\;\;
    \frac{\nu_\eta s_\eta^2 + \sum_{h=1}^H (\theta_{h,t} - \alpha -
    \phi\,\theta_{h,t-1})^2}{\nu_\eta + H}\right).
    \label{eq:gibbs_sigma_eta}
\end{equation}
This is sampled exactly. The long-run mean is recovered as $\mu = \alpha/(1-\phi)$
after each Gibbs iteration.

The AR(1) regression \eqref{eq:ar1_regression} has $H$ observations per
MwG iteration ($H = 55$ in the simulated labour force survey application), from
which two regression parameters $(\alpha, \phi)$ and one variance
$\sigma_\eta^2$ are sampled. The system is therefore well-identified, and
the precision of the AR(1) parameter draws increases with $H$. This is
the identifiability justification for the common-across-strata assumption on
$\phi^{(v)}$ and $\sigma_\eta^{2(v)}$.

 Table~\ref{tab:gibbs} summarises
the sampling method required for each conditional.
\begin{table}[h!]
\centering
\caption{Sampling method for each full conditional in the MwG sampler.}
\label{tab:gibbs}
\begin{tabular}{llll}
\toprule
Group & Conditional & Distribution & Sampling method \\
\midrule
State/link & $\theta_{h,t}^{(v)}$, continuous & Gaussian \eqref{eq:gibbs_theta_cont}
    & Exact Gibbs \\[3pt]
State/link & $\eta_{h,t}^{(v)}$, binary & Non-Gaussian \eqref{eq:gibbs_eta}
    & MwG \\[3pt]
State/link & $\bm{\beta}_t^{(v)}$ & Gaussian \eqref{eq:gibbs_beta}
    & Exact Gibbs \\[3pt]
State/link & $\sigma_u^{2(v)}$ & Inv-$\chi^2$ \eqref{eq:gibbs_sigma_u}
    & Exact Gibbs \\[3pt]
AR(1) & $(\alpha^{(v)}, \phi^{(v)})$ & Truncated Gaussian \eqref{eq:gibbs_alpha_phi}
    & Exact Gibbs via inverse-CDF method \\[3pt]
AR(1) & $\sigma_\eta^{2(v)}$ & Inv-$\chi^2$ \eqref{eq:gibbs_sigma_eta}
    & Exact Gibbs \\[3pt]
AR(1) & $\mu^{(v)}$ & Deterministic & Recovered as $\alpha^{(v)}/(1-\phi^{(v)})$ \\
\bottomrule
\end{tabular}
\end{table}

\section{Sequential HB Update}
\label{sec:b2particle}
The Sequential HB Update (SHBU) is the cross-wave propagation strategy
that carries the HB posterior forward from wave $t-1$ to wave $t$. It
is distinct from the MwG sampler of Section~4, which is the computational
engine within a single MCMC run: one SHBU at wave $t$ propagates the wave $t-1$ posterior to wave $t$
using the Kalman filter approximation described in
Section~\ref{sec:intervals_level}, with the $B$ outputs pooled to
approximate the joint posterior $\pi(\bm{\Theta}_t^{(v)} \mid y_{1:t})$
using \eqref{eq:marginalisation} below. Within each draw $b$, the MwG sampler marginalises over the AR(1) parameters $(\phi^{(v)}, \sigma_\eta^{2(v)}, \mu^{(v)})$ jointly with the stratum states and linking model parameters, so that uncertainty in the AR(1) dynamics is fully propagated into the wave $t$ posterior.
\subsection{The SHBU algorithm}

At wave $t-1$, the MCMC produces $B$ draws from the joint posterior of
$\bm{\Theta}_{t-1}^{(v)}$. The stratum-state draws are retained:
\begin{equation}
    \left\{\hat{\theta}_{h,t-1}^{(v),(b)}\right\}_{b=1}^{B} \;\sim\;
    \pi\!\left(\bm{\theta}_{t-1}^{(v)} \mid y_{1:t-1}\right).
    \label{eq:draws_prev}
\end{equation}
For binary variables these draws are on the logit scale and are non-Gaussian in
general.

The natural approach to sequential Bayesian updating is to summarise
$\pi(\bm{\theta}_{t-1}^{(v)} \mid y_{1:t-1})$ by its posterior mean
and use the AR(1) to propagate this forward as the prior mean at wave
$t$ --- the Kalman filter approach. This is exact for linear Gaussian
models but introduces approximation error for binary variables where
the posterior on the logit scale is non-Gaussian. The implementation
of the SHBU uses this approach for computational feasibility, with
the missing variance from conditioning on the mean rather than the
full posterior recovered analytically via the law of total variance,
as described in Section~\ref{sec:intervals_level}.

For each draw $b = 1,\ldots,B$, the $H$ stratum-state draws
$\{\hat{\theta}_{h,t-1}^{(v),(b)}\}_{h=1}^H$ from wave $t-1$ are treated as
fixed known quantities at wave $t$, entering the AR(1) prior as the
conditioning value. The MCMC for draw $b$ samples the full current-wave parameter
vector $\bm{\Theta}_t^{(v)}$ from the conditional posterior:
\begin{equation}
    \pi\!\left(\bm{\Theta}_t^{(v)} \;\middle|\; y_t,\;
    \left\{\hat{\theta}_{h,t-1}^{(v),(b)}\right\}_{h=1}^H\right).
    \label{eq:cond_post}
\end{equation}
Table~\ref{tab:sampled_fixed} summarises which quantities are sampled and which
are fixed in this conditional posterior. Each parameter marked ``Sampled'' in Table~\ref{tab:sampled_fixed} is
drawn from its full conditional distribution derived in Section~\ref{sec:gibbs};
the correspondence is summarised in Table~\ref{tab:gibbs}.

\begin{table}[h!]
\centering
\caption{Status of each parameter at wave $t$, draw $b$, under SHBU.}
\label{tab:sampled_fixed}
\begin{tabular}{>{\raggedright}p{5.5cm} l p{5cm}}
\toprule
Parameter & Status & Information source \\
\midrule
$\{\theta_{h,t}^{(v)}\}_{h=1}^H$ & \text{Sampled} & Wave-$t$ data $y_t$
    + AR(1) prior \eqref{eq:prop_binary} or \eqref{eq:prop_continuous} \\[4pt]
$\{\theta_{h,t-1}^{(v)}\}_{h=1}^H$ & \text{Fixed} at
    $\hat{\theta}_{h,t-1}^{(v),(b)}$ & Wave $t-1$ MCMC output \\[4pt]
$\bm{\beta}_t^{(v)}$ & \text{Sampled} & Wave-$t$ data $y_t$
    + diffuse prior \eqref{eq:prior_linking}\\[4pt]
$\sigma_u^{2(v)}$ & \text{Sampled} & Wave-$t$ data $y_t$
    + calibrated prior \eqref{eq:prior_linking} \\[4pt]
$\phi^{(v)}$ & \text{Sampled} & $H$ stratum pairs
    $(\hat{\theta}_{h,t-1}^{(v),(b)}, \theta_{h,t}^{(v)})$
    + prior \eqref{eq:prior_ar1} \\[4pt]
$\sigma_\eta^{2(v)}$ & \text{Sampled} & $H$ stratum pairs
    + prior \eqref{eq:prior_ar1} \\[4pt]
$\mu^{(v)}$ & \text{Sampled} & $H$ stratum pairs
    + prior \eqref{eq:prior_ar1} \\
\bottomrule
\end{tabular}
\end{table}

The AR(1) parameters $(\phi^{(v)}, \sigma_\eta^{2(v)}, \mu^{(v)})$ are
identified within each draw $b$ through the $H$ stratum pairs
$(\hat{\theta}_{h,t-1}^{(v),(b)}, \theta_{h,t}^{(v)})$. The fixed
$\hat{\theta}_{h,t-1}^{(v),(b)}$ acts as a regressor in the AR(1) equation, and the
sampled $\theta_{h,t}^{(v)}$ acts as the response. With $H$ strata there are $H$
observations of the AR(1) regression within each draw $b$, from which $\phi^{(v)}$,
$\sigma_\eta^{2(v)}$, and $\mu^{(v)}$ are sampled jointly with the stratum
states and linking model parameters.

\subsection{Validity of the SHBU}

At wave $t$, the update proceeds as follows for each draw $b = 1,\ldots,B$:

\medskip
\noindent\text{Step 1 --- Establish the Bayesian framework for draw $b$.}
Fix the conditioning value $\{\hat{\theta}_{h,t-1}^{(v),(b)}\}_{h=1}^H$ from wave
$t-1$. For each stratum $h$, this determines the structural form of the AR(1)
prior on $\theta_{h,t}^{(v)}$:
\begin{equation}
    \theta_{h,t}^{(v)} \mid \hat{\theta}_{h,t-1}^{(v),(b)},\,
    \phi^{(v)},\, \sigma_\eta^{2(v)},\, \mu^{(v)} \;\sim\;
    \mathcal{N}\!\left(\mu^{(v)} + \phi^{(v)}\!\left(
    \hat{\theta}_{h,t-1}^{(v),(b)} - \mu^{(v)}\right),\; \sigma_\eta^{2(v)}\right),
    \label{eq:ar1_prior}
\end{equation}
where $\hat{\theta}_{h,t-1}^{(v),(b)}$ is fixed and the AR(1) parameters
$(\phi^{(v)}, \sigma_\eta^{2(v)}, \mu^{(v)})$ are unknown quantities to be
sampled in Step~2. Step~1 establishes the full HB model for draw $b$ by
combining the sampling model \eqref{eq:sampling_binary} or
\eqref{eq:sampling_continuous}, the linking model \eqref{eq:linking_binary}
or \eqref{eq:linking_continuous}, and the AR(1) prior \eqref{eq:ar1_prior}
with the priors \eqref{eq:prior_linking} and \eqref{eq:prior_ar1}.

\medskip
\noindent\text{Step 2 --- Run MCMC conditional on draw $b$ from wave $t-1$.}
Sample the full conditional posterior \eqref{eq:cond_post} using the informative AR(1) prior \eqref{eq:ar1_prior}
conditional on the fixed $\hat{\theta}_{h,t-1}^{(v),(b)}$. All parameters in $\bm{\Theta}_t^{(v)}$ are sampled
jointly --- stratum states, linking model parameters, and AR(1) parameters ---
in a single MwG run.

\medskip
\noindent\text{Step 3 --- Aggregate stratum draws to domains.}
\begin{equation}
    \hat{\theta}_{d,t}^{(v),(b)} = \frac{\sum_{h \in \mathcal{H}_d} N_h\,
    \hat{\theta}_{h,t}^{(v),(b)}}{N_d}.
    \label{eq:aggregate}
\end{equation}

A key question is whether pooling the $B$ outputs
$\{\bm{\Theta}_t^{(v),(b)}\}_{b=1}^B$ --- each obtained by conditioning on a
single draw from the wave $t-1$ posterior rather than the full draw set
--- produces a valid sample from the correct joint posterior
$\pi(\bm{\Theta}_t^{(v)} \mid y_{1:t})$.

The answer is yes, by the following marginalisation argument. Each draw $b$
produces a draw from the conditional posterior $\pi(\bm{\Theta}_t^{(v)} \mid
y_t, \hat{\theta}_{h,t-1}^{(v),(b)})$. Since $\hat{\theta}_{h,t-1}^{(v),(b)}$ is itself a
draw from $\pi(\bm{\theta}_{t-1}^{(v)} \mid y_{1:t-1})$, pooling across
$b = 1,\ldots,B$ marginalises over the wave $t-1$ posterior:
\begin{equation}
    \frac{1}{B}\sum_{b=1}^B \pi\!\left(\bm{\Theta}_t^{(v)} \mid y_t,\,
    \hat{\theta}_{h,t-1}^{(v),(b)}\right)
    \;\xrightarrow{B\to\infty}\;
    \int \pi\!\left(\bm{\Theta}_t^{(v)} \mid y_t,\, \bm{\theta}_{t-1}^{(v)}\right)
    \pi\!\left(\bm{\theta}_{t-1}^{(v)} \mid y_{1:t-1}\right)
    \,d\bm{\theta}_{t-1}^{(v)}
    = \pi\!\left(\bm{\Theta}_t^{(v)} \mid y_{1:t}\right).
    \label{eq:marginalisation}
\end{equation}
The convergence in \eqref{eq:marginalisation} follows from the strong
law of large numbers applied to the Monte Carlo average, and the final
equality follows from the Markov property of the state-space model
\eqref{eq:ss_obs}--\eqref{eq:ss_prop}, which implies that $y_t$ and
$y_{1:t-1}$ are conditionally independent given $\bm{\theta}_{t-1}^{(v)}$.
The result is a special case of the sequential Monte Carlo framework of
\citet{gilks2001}, which establishes validity under the more general
setting of weighted particles and multi-wave propagation.

\begin{remark}
The implementation uses the Kalman filter approach: the wave $t-1$
posterior mean is used as the conditioning value for the wave $t$
prior, and the AR(1) propagates this mean forward. This is exact
for continuous variables under a linear Gaussian model. For binary
variables, the posterior on the logit scale is non-Gaussian and the
Gaussian approximation introduces error that accumulates across
waves. The exact SHBU --- conditioning on individual draws rather
than the posterior mean --- would avoid this approximation entirely,
at the cost of $B$ MCMC runs per wave instead of one. The law of
total variance correction in Section~\ref{sec:intervals_level}
recovers the missing between-draw variance exactly for continuous
variables and approximately for binary variables.
\end{remark}

\begin{remark}
The posterior mean $\hat{\theta}_{d,t}^{(v),\mathrm{HB}}$ is the Bayes
estimator of $\theta_{d,t}^{(v)}$ under squared error loss, minimising
the posterior expected squared error over all estimators measurable with
respect to $y_{1:t}$. It uses all historical data $y_1,\ldots,y_t$
automatically through the sequential propagation \eqref{eq:marginalisation}. In classical
repeated surveys, composite estimators of the form
$\hat{\theta}_{d,t}^{\mathrm{comp}} = \alpha_t \hat{\theta}_{d,t}^{\mathrm{direct}}
+ (1-\alpha_t)\hat{\theta}_{d,t-1}^{\mathrm{comp}}$
are constructed to exploit wave-to-wave overlap, with weights $\alpha_t$
chosen to minimise variance subject to approximate design-unbiasedness.  Extending this chain to all previous waves $t=1,\ldots,T$ requires
estimating an increasing number of regression coefficients and
maintaining the full chain of composite estimates, which becomes
operationally intractable as $T$ grows so in practice only the immediately preceding wave is used. 
These are approximations to the Bayes estimator, constructed without a
fully specified model for the temporal dynamics. 
The SHBU posterior mean
is the optimal least-squares estimate of
$\theta_{d,t}^{(v)}$ given $y_{1:t}$ under the stated model, with the
optimality guarantee holding jointly across all strata, domains, and
waves. The price of this optimality is model dependence --- if the AR(1)
propagation model or the linking model are misspecified, the optimality
guarantee no longer holds in a frequentist sense, as the sensitivity
analysis in item 4 of Section~\ref{sec:emp_analysis} confirms.
\end{remark}

\subsection{Level and movement estimators}
\label{sec:estimators}

The following estimators are post-design quantities, computed from
the SHBU posterior draws after wave $t$ data have been collected under the
design $(n_{h,t}^*, m_{h,t}^*)$ determined in Section~\ref{sec:predesign}.

The level estimator at domain $d$, wave $t$, variable $v$ is the posterior mean:
\begin{equation}
    \hat{\theta}_{d,t}^{(v),\mathrm{HB}} = \frac{1}{B}\sum_{b=1}^{B}
    \hat{\theta}_{d,t}^{(v),(b)}.
    \label{eq:level_est}
\end{equation}
The movement estimator uses the same indexed draw $b$ at both waves:
\begin{equation}
    \hat{\Delta}_{d,t}^{(v),\mathrm{HB}} = \frac{1}{B}\sum_{b=1}^{B}
    \left(\hat{\theta}_{d,t}^{(v),(b)} - \hat{\theta}_{d,t-1}^{(v),(b)}\right).
    \label{eq:movement_est}
\end{equation}
For each draw $b$, the wave $t$ estimate $\hat{\theta}_{d,t}^{(v),(b)}$
was generated by conditioning on $\hat{\theta}_{h,t-1}^{(v),(b)}$ as the
AR(1) prior mean and must therefore be differenced against the same
$\hat{\theta}_{d,t-1}^{(v),(b)}$ that it was conditioned on. This pairing
preserves the posterior covariance induced by the AR(1) conditioning and
is the posterior analogue of variance reduction through sample overlap in
classical repeated surveys. The sampling
overlap covariance is captured separately by the sampling variance of movement in Section~\ref{sec:comp1}.

Credible intervals for levels and Calibrated Bayes intervals \citep{little2012} for movements
are defined in Sections~\ref{sec:intervals_level} and~\ref{sec:intervals_movement}, once the sampling and model
variance components have been introduced.

\section{Precision Requirements and the Dynamic Mini-max Strategy}
\label{sec:design}

\subsection{Precision requirements}

The overriding objective of the dynamic mini--max strategy is to meet separate
user-specified precision requirements for levels and movements simultaneously,
across all domains $d = 0,1,\ldots,D$ and all variables $v = 1,\ldots,V$, while
minimising total fieldwork cost. These requirements are specified before the survey
is designed and remain constant across waves.

\medskip
\noindent\textbf{I. Level precision: CV constraint.}
Following \citet{tam2026a}, the level precision requirement is:
\begin{equation}
    \mathrm{CV}\!\left(\hat{\theta}_{d,t}^{(v),\mathrm{HB}}\right) \leq
    g_{\theta,d}^{(v)}, \qquad \forall\, d = 0,1,\ldots,D,\quad v = 1,\ldots,V,
    \label{eq:cv_constraint}
\end{equation}
where $g_{\theta,0}^{(v)} \leq g_{\theta,d}^{(v)}$ for $d \geq 1$ (the national
constraint is at least as tight as the domain constraint). In the simulated labour force survey example,
$g_{\theta,0}^{(v)} = 0.03$ and $g_{\theta,d}^{(v)} = 0.08$.

\medskip
\noindent\textbf{II. Movement precision: power-based SE constraint.}
A CV-based criterion for movement is inappropriate because $\Delta_{d,t}^{(v)}$
may be near zero. The precision requirement for movement is instead grounded in
statistical power. Let $\delta_d^{*(v)}$ denote the minimum detectable
movement for domain $d$ and variable $v$. Let $\alpha$ and $\beta$ denote the
prescribed Type~I and Type~II error rates. The movement precision requirement is:
\begin{equation}
    \widehat{\mathrm{SE}}\!\left(\hat{\Delta}_{d,t}^{(v),\mathrm{HB}}\right) \leq
    g_{\Delta,d}^{(v)}, \qquad g_{\Delta,d}^{(v)} =
    \frac{\delta_d^{*(v)}}{z_{\alpha/2} + z_\beta},
    \label{eq:se_constraint}
\end{equation}
where $z_{\alpha/2}$ and $z_\beta$ are standard normal quantiles. The level and movement precision 
constraints are maintained separately because they have fundamentally different
sensitivity to the design variables: the level constraint is driven primarily by
$n_{h,t}$, while the movement constraint is driven jointly by $n_{h,t}$ and
$m_{h,t}$.  The constraints for the simulated labour force survey example are provided in Section 7.5.4.

\medskip
\subsection{Dynamic Mini-Max Strategy}
\label{sec:threephase}

The Dynamic Mini-Max Strategy requires two inputs before wave $t$ can
be designed: prior calibration data to set the AR(1) hyperparameters,
and SHBU posterior draws $\{\hat{\theta}_{h,t-1}^{(v),(b)}\}_{b=1}^B$
from wave $t-1$ to form $\hat{V}_{\mathrm{mod},d,t}^{(v)}$ and the
conditioning draw set. Two operational scenarios arise depending on whether
historical survey data are available. The transition scenario is
described in Section~\ref{sec:threephase_transition} below; the
cold-start scenario for a new survey is described in
Remark~\ref{rem:coldstart}.

\subsubsection{Transition from an existing classical survey}
\label{sec:threephase_transition}

\noindent\textbf{I. Pre-transition calibration (using $T-1$ historical
waves).}
Before the first mini--max wave, the following quantities are estimated
from the $T-1$ historical waves of classical direct domain estimates
$\{\hat{\theta}_{d,t}^{(v)}\}_{t=1}^{T-1}$:
\begin{itemize}
    \item The pilot AR(1) coefficient $\hat{\phi}^{(v)}$ from regressing
    $\hat{\theta}_{d,t}^{(v)}$ on $\hat{\theta}_{d,t-1}^{(v)}$ across
    waves $t = 2, \ldots, T-1$.
    \item The between-wave correlation $\hat{\rho}_h^{(v)}$ from the
    matched units across historical waves.
    \item The AR(1) innovation scale $\hat{s}_\eta^{2(v)}$ from the
    moment estimator \eqref{eq:moment_scale} with $K = T-1$.
    \item The linking model hyperparameters $(\nu^{(v)}, s^{2(v)})$
    from the four-gate grid search of \citet{tam2026a} applied to the
    most recent wave's data.
\end{itemize}
Using these calibrated priors, the SHBU is run on the most recent
historical wave $t = T-1$ to produce the conditioning draw set
$\{\hat{\theta}_{h,T-1}^{(v),(b)}\}_{b=1}^B$. No additional data
collection is required; the entire calibration uses historical survey
outputs.

\medskip
\noindent\textbf{II. Dynamic mini--max HB strategy --- waves $t \geq T$.}
With the conditioning draw set and calibrated priors in hand, the
Dynamic Mini-Max Design determines $(n_{h,T}^*, m_{h,T}^*)$ at wave
$t = T$. The SHBU then runs at wave $t = T$ and passes the resulting
posterior draws forward to the wave $t=T+1$ design. The strategy continues
wave by wave thereafter.

\begin{remark}
\label{rem:coldstart}
When an NSO launches an entirely new repeated survey with no historical
time series, the treatment of wave $t=2$ depends on the availability
of external calibration data. (i) If AR(1) parameters $(\hat{\phi}^{(v)}, \hat{\sigma}_\eta^{2(v)})$ can
be calibrated from a comparable existing survey programme, they are
used to compute $\hat{V}_{\mathrm{mod},d,2}^{(v)}$ from the wave $t=1$
SHBU draws via \eqref{eq:vmod} below. Wave $t=1$ uses the Bethel allocation of \citet{tam2026a}; from wave
$t=2$ onward the full Dynamic Mini-Max optimisation for
$(n_{h,t}^*, m_{h,t}^*)$ proceeds without modification.  (ii) If no external calibration is available, wave $t=2$ is treated as a
level-only wave: $n_{h,2}^*$ is determined by the static Mini-Max of
\citet{tam2026a} to meet the level CV constraint, and $m_{h,2}^*$ is
chosen via the \citet{patterson1950} optimal overlap formula to
minimise the sampling variance of movement given $n_{h,2}^*$. The
AR(1) parameters are then calibrated from the two-wave series
$\{\hat{\theta}_{d,1}^{(v)}, \hat{\theta}_{d,2}^{(v)}\}$ via
\eqref{eq:moment_scale} with $K=2$, and the full Dynamic Mini-Max
Strategy begins at wave $t=3$. The AR(1) prior calibrated from $K=2$
waves is less precise than in the historical-data case but improves
as further waves accumulate.
\end{remark}

\subsubsection{Dynamic mini--max design}
\label{sec:predesign}

Before wave $t \geq 3$ data are collected, the SE of the movement estimator must
be predicted analytically as a function of the candidate design $(n_{h,t},
m_{h,t})$, in order to check the constraint \eqref{eq:se_constraint}. This
requires decomposing the expected movement variance into a sampling component
(which depends on the design) and a model component (which does not).

\medskip
\noindent\textbf{I. Sampling variance of movement ($V_{\mathrm{samp},d,t}^{(v)}$).}
\label{sec:comp1}
For continuous variable $v$, the sampling variance of the domain mean movement
estimator is:
\begin{equation}
    V_{\mathrm{samp},d,t}^{(v)} = \frac{1}{N_d^2} \sum_{h \in \mathcal{H}_d} N_h^2\,
    \mathrm{DEFF}_h^{(v)} S_h^{2(v)}
    \left[\frac{1-f_{h,t}}{n_{h,t}} + \frac{1-f_{h,t-1}}{n_{h,t-1}}
    - \frac{2\,\rho_h^{(v)}\,m_{h,t}}{n_{h,t}\, n_{h,t-1}}\right],
    \label{eq:vsamp_continuous}
\end{equation}
where $f_{h,t} = n_{h,t}/N_h$, $S_h^{2(v)}$ is the within-stratum variance
assumed stable across waves, and $\rho_h^{(v)} \in [0,1]$ is the between-wave
correlation of the characteristic for matched units in stratum $h$.

For binary variable $v$, The sampling variance for binary variable $v$ is:
\begin{equation}
    V_{\mathrm{samp},d,t}^{(v)} = \frac{1}{N_d^2} \sum_{h \in \mathcal{H}_d}
    \frac{N_h^2}{p_h^{(v)}(1-p_h^{(v)})}
    \left[\frac{1-f_{h,t}^{\mathrm{eff}}}{n_{h,t}^{\mathrm{eff},(v)}}
    + \frac{1-f_{h,t-1}^{\mathrm{eff}}}{n_{h,t-1}^{\mathrm{eff},(v)}}
    - \frac{2\,\rho_h^{(v)}\,m_{h,t}^{\mathrm{eff}}}
    {n_{h,t}^{\mathrm{eff},(v)}\, n_{h,t-1}^{\mathrm{eff},(v)}}\right],
    \label{eq:vsamp_binary}
\end{equation}
where $n_{h,t}^{\mathrm{eff},(v)} = n_{h,t}/\mathrm{DEFF}_h^{(v)}$,
$m_{h,t}^{\mathrm{eff}} = m_{h,t}/\mathrm{DEFF}_h^{(v)}$, and
$f_{h,t}^{\mathrm{eff}} = n_{h,t}^{\mathrm{eff},(v)}/N_h$.
The design effect is fully absorbed into the effective sample sizes.

\begin{remark}
The result for continuous variables follows directly from \citet{patterson1950}
and \citet{cochran1977}.
For binary variables, let $\hat{p}_{h,t}^{(v)} =
r_{h,t}^{(v)}/n_{h,t}^{\mathrm{eff},(v)}$ be the effective-sample proportion.
The sampling covariance between proportions sharing $m_{h,t}^{\mathrm{eff}}$
effective overlapping units is:
\begin{equation}
    \mathrm{Cov}_{\mathrm{samp}}\!\left(\hat{p}_{h,t}^{(v)},
    \hat{p}_{h,t-1}^{(v)}\right) = \frac{m_{h,t}^{\mathrm{eff}}}
    {n_{h,t}^{\mathrm{eff},(v)}\, n_{h,t-1}^{\mathrm{eff},(v)}}\,
    p_h^{(v)}(1 - p_h^{(v)})\,\rho_h^{(v)}.
    \label{eq:cov_binary}
\end{equation}
Applying the delta method to transform from the proportion scale to the logit
scale gives:
\begin{equation}
    \mathrm{Cov}_{\mathrm{samp}}\!\left(\hat{\eta}_{h,t}^{(v)},
    \hat{\eta}_{h,t-1}^{(v)}\right) \approx
    \frac{m_{h,t}^{\mathrm{eff}}}{n_{h,t}^{\mathrm{eff},(v)}\,
    n_{h,t-1}^{\mathrm{eff},(v)}}\,
    \frac{\rho_h^{(v)}}{p_h^{(v)}(1-p_h^{(v)})},
    \label{eq:cov_logit}
\end{equation}
and substituting into the variance of the stratum movement estimator gives
\eqref{eq:vsamp_binary}.
\end{remark}

\medskip
\noindent\textbf{II. Model variance of movement ($V_{\mathrm{mod},d,t}^{(v)}$).}
The model variance is the posterior variance of $\Delta_{d,t}^{(v)}$ arising from
uncertainty in the AR(1) and linking model parameters. It is estimated from the
wave $t-1$ SHBU draws using the law of total variance (Appendix~\ref{app:modvar}):
\begin{equation}
    \hat{V}_{\mathrm{mod},d,t}^{(v)} =
    \frac{1}{B}\sum_{b=1}^{B} \sigma_\eta^{2(v),(b)}
    + \frac{1}{B-1}\sum_{b=1}^{B}
    \left[\tilde{\Delta}_{d,t}^{(v),(b)} -
    \tilde{\Delta}_{d,t}^{(v),\mathrm{HB}}\right]^2,
    \label{eq:vmod}
\end{equation}
where $\tilde{\Delta}_{d,t}^{(v),(b)} = (\phi^{(v),(b)}-1)(\hat{\theta}_{d,t-1}^{(v),(b)}
- \mu^{(v),(b)})$ is the $b$-th draw-specific one-step-ahead movement
prediction --- a deterministic plug-in function of the sampled AR(1)
parameters and the fixed wave $t-1$ state, not a draw from any distribution
---
and $\tilde{\Delta}_{d,t}^{(v),\mathrm{HB}} = \frac{1}{B}\sum_{b=1}^B
\tilde{\Delta}_{d,t}^{(v),(b)}$ is its average. The first term averages the
draw-specific innovation variance — the irreducible movement noise — and the
second term captures the uncertainty in the predicted movement arising from
uncertainty in $\phi^{(v)}$, $\theta_{d,t-1}^{(v)}$, and $\mu^{(v)}$.
This is the irreducible floor on movement variance that the design cannot reduce.

\medskip
\noindent\textbf{III. Pre-design SE estimate.}
The plug-in estimate of the total movement SE for a candidate design is:
\begin{equation}
    \widehat{\mathrm{SE}}\!\left(\hat{\Delta}_{d,t}^{(v),\mathrm{HB}}\right) =
    \sqrt{V_{\mathrm{samp},d,t}^{(v)}\!\left(n_{h,t}, m_{h,t}\right) +
    \hat{V}_{\mathrm{mod},d,t}^{(v)}},
    \label{eq:se_est}
\end{equation}
where $V_{\mathrm{samp},d,t}^{(v)}$ is computed analytically for the candidate $(n_{h,t}, m_{h,t})$, and
$\hat{V}_{\mathrm{mod},d,t}^{(v)}$ is fixed from the wave $t-1$ SHBU output.
This expression makes the dependence of movement precision on both design
variables explicit and enables the optimisation.

\subsubsection{Level interval}
\label{sec:intervals_level}

Once the optimal design $(n_{h,t}^*, m_{h,t}^*)$ has been implemented
and the SHBU posterior draws $\{\hat{\theta}_{d,t}^{(v),(b)}\}_{b=1}^B$
obtained, the level interval is published.

Under the exact SHBU --- running $B$ separate MCMC chains, one
conditioned on each draw $\hat{\theta}_{h,t-1}^{(b)}$ from the wave
$t-1$ posterior --- the $B$ pooled draws give a genuine posterior
credible interval with no further adjustment:
\begin{equation}
    \mathrm{CI}_{95\%}^{(v)}\!\left(\theta_{d,t}^{(v)}\right)
    = \left[Q_{0.025}\!\left(\hat{\theta}_{d,t}^{(v),(b)}\right),\;
    Q_{0.975}\!\left(\hat{\theta}_{d,t}^{(v),(b)}\right)\right],
    \label{eq:ci_level}
\end{equation}
where $Q_p$ denotes the $p$-th empirical quantile over the $B$ draws.
This credible interval reflects uncertainty in $\theta_{d,t}^{(v)}$
arising from the sampling model, the linking model, and the AR(1)
propagation model jointly, given all data $y_{1:t}$. No additional
sampling variance adjustment is required since the sampling uncertainty
is already incorporated in the posterior draws through the likelihood - 
for continuous variables, the sampling variance $\psi_{h,t}^{(v)}$
enters directly through the Fay--Herriot sampling model
\eqref{eq:sampling_continuous}; for binary variables, the design
effect is absorbed into the effective sample size
$n_{h,t}^{\mathrm{eff},(v)}$ in the Binomial likelihood
\eqref{eq:sampling_binary}.

In practice, the exact SHBU is computationally prohibitive for large
$B$ and $M$. The implementation uses the wave $t-1$ posterior mean
$\bar{\theta}_{h,t-1} = \frac{1}{B}\sum_{b=1}^B
\hat{\theta}_{h,t-1}^{(b)}$ as a single fixed conditioning value
for a single \texttt{mcmcsae} chain at wave $t$. Conditioning on the
mean rather than on the full distribution omits the between-draw
variance --- the uncertainty in $\theta_{t-1}$ itself. By the law of
total variance this missing term is $\hat{\phi}^{(v)2}
\hat{V}_{\mathrm{HB},t-1}^{(v)}$, giving:
\begin{equation}
  \mathrm{CBI}_{95\%}^{(v)}\!\left(\theta_{d,t}^{(v)}\right)
  = \hat{\theta}_{d,t}^{(v),\mathrm{HB}} \;\pm\;
  z_{0.975}\sqrt{\hat{V}_{\mathrm{HB},t}^{(v)} +
  \hat{\phi}^{(v)2}\,\hat{V}_{\mathrm{HB},t-1}^{(v)}},
  \label{eq:cbi_level_impl}
\end{equation}
where $\hat{V}_{\mathrm{HB},t}^{(v)}$ is the posterior variance from
the single chain and $\hat{V}_{\mathrm{HB},t-1}^{(v)}$ is the wave
$t-1$ \emph{corrected} CBI variance --- itself the result of applying
the same law-of-total-variance correction at wave $t-1$. Through this
recursion, the correction accumulates the full geometric series
$\sum_{k=1}^{t-1}\hat{\phi}^{(v)2k}\hat{V}_{\mathrm{chain},t-k}^{(v)}$
across all previous waves, converging rapidly since
$|\hat{\phi}^{(v)}| < 1$. This correction is exact under the Gaussian
linear model for continuous variables; for binary variables it is
approximate. Because the interval combines posterior draws with an
analytically added component it is a Calibrated Bayes interval (CBI)
in the spirit of \citet{little2012} rather than a pure posterior
credible interval.

\subsubsection{Movement interval}
\label{sec:intervals_movement}

For movement, theory and computational implementation coincide: the
movement interval is a Calibrated Bayes interval by construction,
regardless of whether the exact or approximate SHBU is used. This
is because the posterior draws from the SHBU capture only the model
variance component $\hat{V}_{\mathrm{mod},d,t}^{(v)}$; the sampling
variance $V_{\mathrm{samp},d,t}^{(v)}$ arises from the survey design
and must always be incorporated explicitly.

The movement estimator $\hat{\Delta}_{d,t}^{(v),\mathrm{HB}}$ has
total variance comprising both the model component
$\hat{V}_{\mathrm{mod},d,t}^{(v)}$ --- captured by the spread of
the posterior draws --- and the sampling component
$V_{\mathrm{samp},d,t}^{(v)}$ --- arising from the survey design.
Under the approximate normality of
$\hat{\Delta}_{d,t}^{(v),\mathrm{HB}}$ --- which holds for
continuous variables by the linearity of the posterior given the
AR(1) parameters, and approximately for binary variables when domain
sample sizes are moderate to large --- the movement CBI,
constructed in the spirit of calibrated Bayes \citep{little2012},
is:
\begin{equation}
    \mathrm{CBI}_{95\%}^{(v)}\!\left(\Delta_{d,t}^{(v)}\right)
    = \hat{\Delta}_{d,t}^{(v),\mathrm{HB}} \;\pm\;
    z_{0.975}\,\widehat{\mathrm{SE}}\!\left(\hat{\Delta}_{d,t}^{(v),
    \mathrm{HB}}\right),
    \label{eq:cbi_movement}
\end{equation}
where $\widehat{\mathrm{SE}}(\hat{\Delta}_{d,t}^{(v),\mathrm{HB}}) =
\sqrt{V_{\mathrm{samp},d,t}^{(v)}(n_{h,t}^*, m_{h,t}^*) +
\hat{V}_{\mathrm{mod},d,t}^{(v)}}$ from \eqref{eq:se_est} and
$z_{0.975} = 1.960$. The Dynamic Mini-Max Design ensures
$\widehat{\mathrm{SE}} \leq g_{\Delta,d}^{(v)}$ by construction,
so that the minimum detectable movement $\delta_d^{*(v)} =
2.802\,g_{\Delta,d}^{(v)}$ is achievable at the prescribed power
under the stated model. The interval \eqref{eq:cbi_movement} is
designed to achieve approximately $95\%$ frequentist coverage when
the model is correctly specified; empirical verification across
$M = 200$ Monte Carlo replications is reported in
Section~\ref{sec:emp_results}. The level CBI
\eqref{eq:cbi_level_impl} and movement CBI \eqref{eq:cbi_movement}
are the two publishable inferential bands from the Sequential HB
Estimation at each wave.

\subsubsection{The dynamic mini--max optimisation}

The Dynamic Mini-Max Design at each wave $t \geq 3$ finds the
smallest cost design $(n_{h,t}, m_{h,t})$ that simultaneously
satisfies all user-specified precision constraints for levels and
movements across all domains and variables, and subject to a maximum overlap fraction $\rho^{\max}$ per stratum and a fieldwork cost budget $B^{\max}$. Given the HB state-space
model, the solution to this problem is the smallest cost design that
jointly meets both constraint sets simultaneously --- the dynamic
analogue of the mini-max result of \citet{tam2026a}. Formally:
\begin{equation}
    \min_{\{n_{h,t},\, m_{h,t}\}} C_t\!\left(\{n_{h,t}\}, \{m_{h,t}\}\right)
    \label{eq:opt_obj}
\end{equation}
subject to:
\begin{align}
    \mathrm{CV}\!\left(\hat{\theta}_{d,t}^{(v),\mathrm{HB}}\right)
    &\leq g_{\theta,d}^{(v)},
    \quad \forall\, d,\; v,
    \label{eq:opt_cv}\\
    V_{\mathrm{samp},d,t}^{(v)}\!\left(n_{h,t}, m_{h,t}\right) + \hat{V}_{\mathrm{mod},d,t}^{(v)}
    &\leq \left[g_{\Delta,d}^{(v)}\right]^2,
    \quad \forall\, d,\; v,
    \label{eq:opt_se}\\
    0 \leq m_{h,t} &\leq \min\!\left(n_{h,t-1},\, n_{h,t}\right),
    \quad \forall\, h,
    \label{eq:opt_overlap}\\
    n_{h,t} &\geq n_{h}^{\min},\quad n_{h,t} \in \mathbb{Z}^+,
    \quad \forall\, h,
    \label{eq:opt_min}\\
    m_{h,t} &\leq \lfloor \rho^{\max} \cdot n_{h,t} \rfloor,
    \quad \forall\, h,
    \label{eq:opt_maxoverlap}\\
    C_t\!\left(\{n_{h,t}\}, \{m_{h,t}\}\right) &\leq B^{\max},
    \label{eq:opt_budget}
\end{align}
where the total fieldwork cost is:
\begin{equation}
    C_t\!\left(\{n_{h,t}\}, \{m_{h,t}\}\right) = \sum_{h=1}^{H}
    \left[c_h^{\mathrm{new}}\,(n_{h,t} - m_{h,t}) + c_h^{\mathrm{ret}}\,
    m_{h,t}\right].
    \label{eq:cost}
\end{equation}
Since $c_h^{\mathrm{new}} \geq c_h^{\mathrm{ret}}$, the cost is non-increasing
in $m_{h,t}$ for fixed $n_{h,t}$: there is a direct cost incentive to retain
units, reinforcing the precision incentive from the movement constraint.

Constraint \eqref{eq:opt_maxoverlap} imposes a maximum overlap fraction
$\rho^{\max} \in (0,1)$ per stratum. Without this bound the optimiser
would retain all available units ($m_{h,t} = \min(n_{h,t-1}, n_{h,t})$),
eliminating fresh recruitment and exposing the survey to bias at the next wave. In practice $\rho^{\max}$ and $B^{\max}$ are set
by the NSO to reflect respondent burden and affordability considerations respectively.

The movement constraint \eqref{eq:opt_se} implies a lower bound on the
precision target $g_{\Delta,d}^{(v)}$ that the NSO can feasibly specify.
Rearranging \eqref{eq:opt_se}, the constraint requires:
\begin{equation}
    \left[g_{\Delta,d}^{(v)}\right]^2 \geq
    V_{\mathrm{samp},d,t}^{(v)}\!\left(n_{h,t}, m_{h,t}\right)
    + \hat{V}_{\mathrm{mod},d,t}^{(v)}.
    \label{eq:se_residual}
\end{equation}
Since $V_{\mathrm{samp},d,t}^{(v)} \geq V_{\mathrm{samp},d,t}^{(v),\min}
> 0$ for any finite sample, and $V_{\mathrm{mod},d,t}^{(v)} > 0$
regardless of sample size, the right side of \eqref{eq:se_residual}
is bounded away from zero. The NSO cannot set $g_{\Delta,d}^{(v)}$
below $\sqrt{V_{\mathrm{samp},d,t}^{(v),\min} +
V_{\mathrm{mod},d,t}^{(v)}}$ and expect the constraint to be
feasible --- this is the crucial input to the infeasibility floor: the NSO cannot set $g_{\Delta,d}^{(v)}$ below $\sqrt{V_{\mathrm{samp},d,t}^{(v),\min} + \hat{V}_{\mathrm{mod},d,t}^{(v)}}$.

\begin{remark}
The four sets of constraints in the optimisation
\eqref{eq:opt_obj}--\eqref{eq:opt_budget} interact in a structured way
that is worth making explicit.
\begin{enumerate}
\item The CV constraint \eqref{eq:opt_cv} determines the minimum sample
size $n_{h,t}$ required to achieve the desired level precision. Once
satisfied, $n_{h,t}$ is effectively fixed: reducing $n_{h,t}$ below
this floor violates the CV target.
\item With $n_{h,t}$ fixed, the SE constraint \eqref{eq:opt_se} places
a lower bound on $m_{h,t}$: sufficient overlap is needed to
reduce $V_{\mathrm{samp},d,t}^{(v)}$ so that the movement SE target
is met. The $\hat{V}_{\mathrm{mod},d,t}^{(v)}$ component is not reduced by sample or overlap size.
\item The maximum overlap constraint \eqref{eq:opt_maxoverlap} places
an upper bound on $m_{h,t}$ to prevent bias at the next wave.
\item The budget constraint \eqref{eq:opt_budget}, since cost
\eqref{eq:cost} is decreasing in $m_{h,t}$ for fixed $n_{h,t}$, places
a lower bound on $m_{h,t}$: a tighter budget requires more
overlap. 
\end{enumerate}
The feasible set for $m_{h,t}$ is therefore the interval
$[m_{h,t}^{\mathrm{lb}}, m_{h,t}^{\mathrm{ub}}]$ where the lower
bound is the maximum of the SE-driven and budget-driven floors, and the
upper bound is the minimum of $\min(n_{h,t-1}, n_{h,t})$ and the
$\rho^{\max}$ ceiling. Feasibility requires this interval to be
non-empty; infeasibility arises when the SE target is so tight that the
required $m_{h,t}$ exceeds the $\rho^{\max}$ cap, or when the budget
is so tight that the required $m_{h,t}$ exceeds $\min(n_{h,t-1},
n_{h,t})$.
\end{remark}

\begin{remark}
The optimisation \eqref{eq:opt_obj}--\eqref{eq:opt_budget} is a
constrained mixed-integer programme. The algorithm implemented here
uses a sequential stratum-by-stratum search and is not guaranteed to
find the global optimum; it finds a locally optimal solution that
satisfies all constraints. This is standard practice in survey
allocation algorithms such as Bethel \citep{bethel1989}, which also
relies on iterative search without a global optimality guarantee.
\end{remark}

\subsubsection{Sequential HB estimation}
\label{sec:postdesign}

With the optimal design $(n_{h,t}^*, m_{h,t}^*)$ from the Dynamic Mini-Max
Design, the wave $t$ survey is fielded and data collected. The SHBU of
Section~\ref{sec:b2particle} is then run with sampling variance
$\psi_{h,t}^{(v)}$ and effective sample size $n_{h,t}^{\mathrm{eff},(v)}$
determined by $(n_{h,t}^*, m_{h,t}^*)$, producing $B$ posterior draws
$\{\hat{\theta}_{d,t}^{(v),(b)}\}_{b=1}^B$. Level and movement estimates and
credible intervals follow directly from Section~\ref{sec:estimators} and
Sections~\ref{sec:intervals_level} and~\ref{sec:intervals_movement}, with no approximation: the posterior draws
capture the full uncertainty in both levels and movements, including
uncertainty in the AR(1) and linking model parameters.

The output of the Sequential HB Estimation at wave $t$ feeds forward into
the Dynamic Mini-Max Design at wave $t+1$ through the retained stratum-state
draws $\{\hat{\theta}_{h,t}^{(v),(b)}\}_{b=1}^B$ and AR(1) parameter draws
$\{(\phi^{(v),(b)}, \sigma_\eta^{2(v),(b)})\}_{b=1}^B$. The connection is
one-directional within each wave: the design determines the data collection,
which determines the posterior, which informs the next design. The
wave-by-wave procedure then repeats, with the pre-transition calibration of
Section~\ref{sec:threephase} performed once before the strategy begins.

\section{Simulation Study Design and Results}
\label{sec:simulation}

The numerical example uses a four-wave simulation designed to illustrate the
Dynamic Mini-Max Design and Sequential HB Estimation on data that are
structurally consistent with the Australian monthly labour force statistics.
The simulation has two components: a real microdata set at wave $t=1$ from
the 2021 Australian Census, and simulated data sets for waves
$t=2,3,4$ based on the AR(1) model calibrated to historical ABS monthly Labour Force Survey ((LFS) data. The AR(1) parameters used for data generation are calibrated once from historical ABS LFS data and treated as fixed known truths throughout the simulation, i.e. all M=200 replications share the same data-generating dynamics. These fixed truths are distinct from the posterior draws of the same parameters produced by the MwG sampler, which vary across replications as they are estimated anew from each replication's simulated survey data.  The Bethel allocation was performed using R2BEAT \citep{falorsi2021} and MCMC sampling performed using the mcmcsae package  \citep{boonstra2021}, including the Metropolis-within-Gibbs sampler for Conditional 2 in Section 4 above.

\subsection{Wave $t=1$: Australian Census microdata}

The wave $t=1$ population is the 2021 Australian Census microdata, following
\citet{tam2026b}. The population comprises $N = 840{,}402$ working-age
individuals across $H=55$ geographic strata linked to $D = 8$ state and territory
domains plus the national domain. Three target variables are defined as
per-person means over the civilian population aged 15 and over:

\begin{itemize}
    \item \textbf{Employment} (binary): $y_{i,1}^{(\text{emp})} = 1$ if person
    $i$ is employed, 0 otherwise. The stratum mean $\theta_{h,1}^{(\text{emp})}$
    is the employment-to-population ratio.
    \item \textbf{Unemployment} (binary): $y_{i,1}^{(\text{unemp})} = 1$ if
    person $i$ is unemployed, 0 otherwise. The stratum mean
    $\theta_{h,1}^{(\text{unemp})}$ is the unemployment-to-population ratio.
    \item \textbf{Hours worked} (continuous): $y_{i,1}^{(\text{hrs})} = h_{i,1}$,
    the weekly hours worked by person $i$ including zeros for non-employed
    persons. The stratum mean $\theta_{h,1}^{(\text{hrs})}$ is mean per capita weekly hours.
\end{itemize}

True stratum population means $\theta_{h,1}^{(v),\mathrm{true}}$ are computed
directly from the Census microdata. True domain means follow from
\eqref{eq:domain_agg}. The wave $t=1$ cross-sectional HB model of
\citet{tam2026a} is applied to produce the starting draw set
$\{\theta_{h,1}^{(v),(b)}\}_{b=1}^B$, which initiates the SHBU for wave $t=2$.

\subsection{Waves $t=2,3,4$}

To generate waves $t=2,3,4$ in a manner consistent with real monthly LFS
dynamics, the AR(1) parameters $(\phi^{(v)}, \sigma_\eta^{2(v)}, \mu^{(v)})$
are calibrated from ABS monthly LFS state-level estimates. For each variable
$v$, the state-level per-person means $\{\hat{\theta}_{d,k}^{(v)}\}$ are
computed from published ABS counts and civilian population denominators over
$K=68$ historical months from January 2016 to August 2021. The calibration proceeds as follows:

\begin{enumerate}
    \item \textbf{Long-run mean:} $\hat{\mu}^{(v)} = \frac{1}{K \times D}
    \sum_{d=1}^D \sum_{k=1}^K \hat{\theta}_{d,k}^{(v)}$, the grand mean across
    states and months.

    \item \textbf{AR(1) coefficient:} $\hat{\phi}^{(v)}$ from a pooled AR(1)
    regression of $\hat{\theta}_{d,k}^{(v)}$ on $\hat{\theta}_{d,k-1}^{(v)}$
    across all states $d$ and months $k=2,\ldots,K$, with a common slope
    consistent with the common-$\phi$ assumption of the model.

    \item \textbf{Innovation variance:} $\hat{\sigma}_\eta^{2(v)}$ from the
    moment estimator \eqref{eq:moment_scale} with $K$ historical waves,
    corrected for the 7/8 monthly overlap of the ABS LFS design via
    \eqref{eq:psi_corr}.

    \item \textbf{Between-wave correlation:} $\hat{\rho}_h^{(v)}$ estimated
    from the $7n_{h,t}/8$ matched units in the historical LFS.
\end{enumerate}

These calibrated values $(\hat{\phi}^{(v)}, \hat{\sigma}_\eta^{2(v)},
\hat{\mu}^{(v)}, \hat{\rho}_h^{(v)})$ serve as the true
data-generating parameters for the simulation and appear directly in
the four-step population generation procedure below. They are held fixed
across all waves.

The synthetic microdata populations at waves $t = 2, 3, 4$ are
generated by applying
the following four steps to the wave $t=1$ Census microdata. 

\medskip
\textbf{Step 1 --- AR(1) propagation of target states.}
Target stratum means $\tilde{\theta}_{h,t}^{(v)}$ are propagated
from the previous wave using the calibrated AR(1) parameters:
\begin{equation}
  \tilde{\theta}_{h,t}^{(v)} = \hat{\mu}^{(v)} +
  \hat{\phi}^{(v)}\!\left(\theta_{h,t-1}^{(v),\mathrm{true}} -
  \hat{\mu}^{(v)}\right) + \eta_{h,t}^{(v)},
  \qquad
  \eta_{h,t}^{(v)} \overset{\mathrm{iid}}{\sim}
  \mathcal{N}\!\left(0,\,\hat{\sigma}_\eta^{2(v)}\right),
  \label{eq:sim_dgp}
\end{equation}
where the innovations $\eta_{h,t}^{(v)}$ are drawn independently
for each stratum and wave \emph{once} as part of the population
generation and are thereafter fixed constants; they do not vary
across the $M = 200$ Monte Carlo replications. For binary variables,
\eqref{eq:sim_dgp} operates on the logit scale and
$\tilde{p}_{h,t}^{(v)} = \mathrm{logistic}(\tilde{\theta}_{h,t}^{(v)})$
is the implied target proportion. These target states govern the
microdata update in Step~2.

\medskip
\textbf{Step 2 --- Open-population update (exits and entries).}
The wave $t-1$ microdata population is updated through an entry-exit
mechanism. A fraction $r_{\mathrm{exit}} = 0.8\%$ of existing records
is randomly removed from each stratum (deaths and emigration). A
fraction $r_{\mathrm{entry}} \approx 0.965\%$ of new individual
records is then added, with outcomes drawn using the target states
from Step~1:
\begin{itemize}
\item \textit{Binary variables (new entrants):}
  $y_{i,t}^{(v)} \sim \mathrm{Bernoulli}(\tilde{p}_{h,t}^{(v)})$;
\item \textit{Hours Worked (new entrants):}
  $y_{i,t}^{(\mathrm{hrs})} = \max\!\left(0,\;
  \mathcal{N}(\tilde{\theta}_{h,t}^{(\mathrm{hrs})},\,
  5\hat{\sigma}_\eta^{2(\mathrm{hrs})})\right)$,
  where negative draws are set zero. The variance
  is scaled up by a factor of 5 to approximate the wider
  cross-sectional spread. Since published ABS hours worked data are
  unavailable for direct calibration, the AR(1) parameters for
  Hours Worked are assumed values rather than empirically
  calibrated quantities.
\end{itemize}
For retained units carried forward from wave $t-1$, the calibrated
between-wave correlation $\hat{\rho}_h^{(v)}$ governs the update:
\begin{itemize}
\item \textit{Binary variables (retained units):} each unit's
  outcome is flipped with probability $1 - \hat{\rho}_h^{(v)}$
  and retained with probability $\hat{\rho}_h^{(v)}$, inducing
  the target between-wave correlation at the stratum level;
\item \textit{Hours Worked (retained units):} an individual-level
  AR(1) update is applied using the calibrated parameters
  $(\hat{\phi}^{(\mathrm{hrs})}, \hat{\sigma}_\eta^{2(\mathrm{hrs})},
  \hat{\mu}^{(\mathrm{hrs})})$:
  \[
    y_{i,t}^{(\mathrm{hrs})} = \hat{\mu}^{(\mathrm{hrs})} +
    \hat{\phi}^{(\mathrm{hrs})}\!\left(y_{i,t-1}^{(\mathrm{hrs})}
    - \hat{\mu}^{(\mathrm{hrs})}\right) + \varepsilon_{i,t},
    \qquad
    \varepsilon_{i,t} \sim
    \mathcal{N}(0,\,\hat{\sigma}_\eta^{2(\mathrm{hrs})}).
  \]
\end{itemize}

\medskip
\textbf{Step 3 --- Recompute true stratum means.}
After all exits, entries, and retained-unit updates have been applied,
the true stratum mean is recomputed as the population mean over all
individual records in stratum $h$ at wave $t$:
\[
  \theta_{h,t}^{(v),\mathrm{true}} =
  \frac{1}{N_{h,t}} \sum_{i \in U_{h,t}} y_{i,t}^{(v)},
\]
where $U_{h,t}$ is the updated stratum population of size $N_{h,t}$.
Because of the stochastic entry-exit process,
$\theta_{h,t}^{(v),\mathrm{true}}$ generally differs slightly from
the AR(1) target $\tilde{\theta}_{h,t}^{(v)}$ from Step~1. The
recomputed mean is the population truth used in all Monte Carlo
evaluations.

\medskip
\textbf{Step 4 --- Store the real (wave 1) and synthetic populations (waves 2 to 4).}
The updated microdata population and recomputed truth values are
stored as fixed files. These populations are generated \emph{once}
and held constant; the $M = 200$ Monte Carlo replications consist
solely of repeated stratified sampling from these same fixed
populations, not regeneration of the populations.

\subsection{Simulating survey samples at each Monte Carlo replication}

At each replication $m$ and wave $t$, a stratified random sample of size
$n_{h,t}$ is drawn without replacement from the fixed synthetic population
at wave $t$ --- the Census microdata at $t=1$ and the simulated populations
at $t=2,3,4$. Matched units from wave $t-1$ are
identified by their unique person identifier and retained in the wave $t$
sample up to the overlap target $m_{h,t}$; their wave $t$ outcome values
are read directly from the wave $t$ population file. Stratum direct
estimates $\bar{y}_{h,t}^{(v),(m)}$ are computed from the sample as simple
means.

Two scenarios are compared at each replication:

\medskip
\begin{enumerate}

\item \textbf{Scenario A --- Fixed sample size and rotation:}
Sample size $n_{h,t}^{A} = 0.05 \times N_{h,1}$,
a proportional allocation at the Census survey fraction of 5\%, held constant
across waves. This gives a total classical sample of $n^{A} = 42{,}018$.
This allocation serves as the common starting point for both designs:
analytically, it largely meets the level CV precision constraints across
all domains, with the only marginal exceptions being NT and ACT
Unemployment (achieved CV $\approx 27\%$ against a 25\% target),
which would require only a trivial Bethel reallocation to correct.
Overlap is set to the 7/8 rotation used by the Australian Bureau of Statistics:
$m_{h,t}^{A} = \lfloor 7n_{h,t}^{A}/8 \rfloor$.
Direct estimates are computed at each wave and used to evaluate cost,
CV precision and SE feasibility targets.

\item \textbf{Scenario B --- Optimal overlap ratio:}
Under Scenario~B (Dynamic Mini-Max), at each replication and each wave
$t\geq3$, the Dynamic Mini-Max Design is solved using the wave $t-1$
SHBU draws to determine
$(n_{h,t}^{*,(m)},m_{h,t}^{*,(m)})$.
Cost parameters:
$c_h^{\mathrm{new}}=2$,
$c_h^{\mathrm{ret}}=1$
for all $h$.
Overlap and budget constraints:
$\rho^{\max} = 9/10$ (maximum fraction of each stratum
that may be retained, enforcing at least $1/9$ fresh recruitment per
wave to limit  bias at the next wave); and $B^{\max} = 45{,}000$ cost
units.
Movement precision targets, $\delta_d^{*(v)}$, set to values consistent with the calibrated
$\hat{\sigma}_{\eta}^{2(v)}$,
$\alpha=0.05$,
$\beta=0.20$, and level targets,
$g_{\theta,0}^{(v)}$, and
$g_{\theta,d}^{(v)}$, are provided in Table~\ref{tab:constraints} below.  The optimal overlap ratio is
$m_{h,t}^{*,(m)}/n_{h,t}^{*,(m)}$
under Scenario~B, compared with the classical $7/8$.

\end{enumerate}

\subsection{Monte Carlo evaluation}

The study runs $M = 200$ replications. At each replication, the SHBU is run
with $B = 2{,}000$ post-burn-in MCMC draws and a burn-in of 500 iterations,
using the \texttt{mcmcsae} package \citep{boonstra2021}.

\begin{remark}[Implementation of the SHBU]
As described in Sections~\ref{sec:intervals_level} and~\ref{sec:intervals_movement}, the
implementation uses the wave $t-1$ posterior mean as a single fixed
conditioning value rather than running $B$ separate MCMC chains. The
missing variance is recovered analytically via the law of total
variance, giving Calibrated Bayes intervals (CBIs) rather than pure
posterior credible intervals for both levels and movements. This is
why results in this section are reported as CBIs. The $M = 200$
simulation confirms that both CBIs achieve the nominal coverage
properties reported below.
\end{remark}

The following quantities are recorded at each replication $m$, wave $t$,
domain $d$, variable $v$, and scenario $s$:

\begin{enumerate}
    \item \textbf{Level coverage:} whether $\theta_{d,t}^{(v),\mathrm{true}}$
    falls within the 95\% Calibrated Bayes interval (CBI) for the level.
    \item \textbf{Movement coverage:} whether $\Delta_{d,t}^{(v),\mathrm{true}}$
    falls within the 95\% Calibrated Bayes interval (CBI) for the movement.
    \item \textbf{Total cost:} $C_t^{(m)} = \sum_h [c_h^{\mathrm{new}}
    (n_{h,t} - m_{h,t}) + c_h^{\mathrm{ret}} m_{h,t}]$.
    \item \textbf{Optimal overlap ratio:} $m_{h,t}^{*,(m)}/n_{h,t}^{*,(m)}$
    under Scenario~B, compared to the classical 7/8.
\end{enumerate}

These quantities are reported in the next Section.

\subsection{Results and Analysis}
\label{sec:results}

\subsubsection{Results}
\label{sec:emp_results}

The detailed level CV and movement SE
targets to be met by the estimators of the empirical example are provided in Table~\ref{tab:constraints} below. In the sequel, we group the 8 States and Territories of Australia into large, medium and small domains.  \text{National} comprises all States and Territories combined.  The large domain comprises NSW, VIC, QLD, WA, while the medium and small domains comprises SA, TAS; and ACT, NT respectively. 

\begin{table}[htbp]
\centering
\caption{Precision targets by domain class and variable -
$g_{\theta,d}^{(v)}$ (\%) and 
$\delta_d^{*(v)}$
(Employment/Unemployment in percentage points (pp); Hours Worked in hrs/week).}
\label{tab:constraints}
\small
\begin{tabular}{lcccccc}
\hline
 & \multicolumn{3}{c}{Level CV target (\%)} & \multicolumn{3}{c}{Movement target $\delta^*$} \\
\cmidrule(lr){2-4}\cmidrule(lr){5-7}
Domain class & Empl & Unemp & Hrs & Empl (pp) & Unemp (pp) & Hrs (hrs) \\
\hline
National                    & 3  & 5  & 5  & 3.0 & 2.0 & 6.0 \\
Large (NSW, VIC, QLD, WA)  & 5  & 10 & 8  & 5.0 & 5.0 & 6.0 \\
Medium (SA, TAS)            & 8  & 20 & 12 & 11.0 & 22.0 & 7.0 \\
Small (ACT, NT)             & 10 & 25 & 15 & 25.0 & 60.0 & 8.0 \\
\hline
\end{tabular}
\end{table}

The movement targets $\delta_d^{*(v)}$ in Table~\ref{tab:constraints}
are chosen to be feasible given the model variance
$\hat{V}_{\mathrm{mod},d,t}^{(v)}$ and the planned sample size,
rather than to reflect operationally realistic movement magnitudes.
At the domain level, movements of the stated magnitude are rarely
if ever observed in Australian labour force data; the targets are
correspondingly loose and are not the binding constraints in the
optimisation for most domains. The binding constraints are the
level CV targets, particularly for small domains. In an operational
setting an NSO would specify tighter movement targets --- consistent
with historically observed movements of 1--2pp nationally for
employment and unemployment --- which would require a larger sample
size and would make the movement SE constraint genuinely binding
across all domains. The present simulation uses loose targets to
demonstrate the feasibility of the DMM framework across all domain
classes, including small domains where tight movement targets are
statistically infeasible at any practical sample size.
Detailed simulation results are provided in the Supplementary
Materials.

It should be noted that the movement targets for Employment and
Unemployment are expressed in percentage points of the
unemployment-to-population ratio and the employment-to-population
ratio respectively --- the proportion of all civilians aged 15 and
over who are unemployed or employed --- rather than the conventional
unemployment rate (unemployed persons as a proportion of the labour
force). Since the unemployment-to-population ratio $\theta^{(v),\mathrm{true}}$
is approximately 0.034 in the simulation, movement targets such as
22pp for Medium domains would imply a shift from roughly 3\% to 25\%
--- operationally unrealistic but set at a level that is statistically
feasible given $\hat{V}_{\mathrm{mod}}$ at the planned sample size.

\begin{figure}[htbp]
\centering
\centering\input{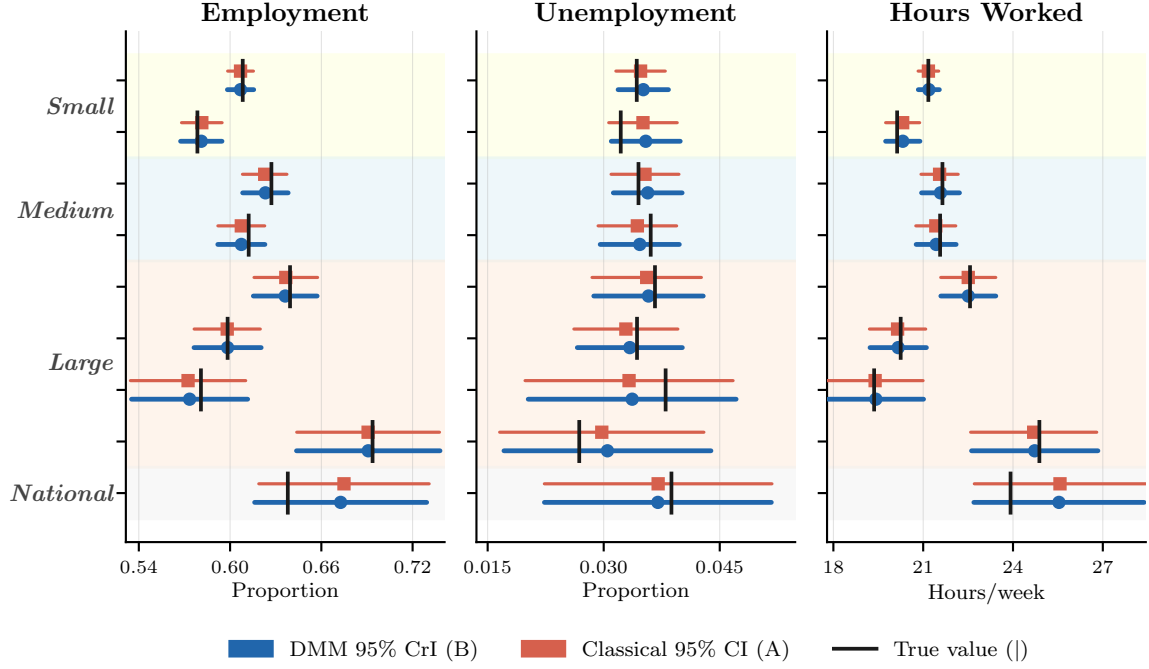}
\caption{Mean level estimates and 95\% CBI/confidence intervals at
wave $t=4$ across all nine domains for Employment, Unemployment and Hours Worked
($M=200$ replications). Blue circles (\textbullet): DMM/SHBU (Scenario~B).
Red squares ($\blacksquare$): classical 7/8 rotation (Scenario~A).
Black bar ($|$): true population value.
Background shading: National (white), Large (light orange) = NSW/VIC/QLD/WA,
Medium (light blue) = SA/TAS, Small (light yellow) = ACT/NT.}
\label{fig:level}
\end{figure}

\paragraph{1.\ Level estimates (Figure~\ref{fig:level}).}
Both designs begin from the same 5\% proportional allocation ($n^A = 42{,}018$).
Analytically, this allocation meets the level CV constraints for all domains
and variables except NT and ACT Unemployment, where the achieved CV
of approximately 27\% marginally exceeds the 25\% target.  A classical
Bethel reallocation would correct this with a negligible increase in total
sample size.  The DMM, by exploiting HB borrowing of strength, meets
the NT and ACT Unemployment CV constraints at $n^* = 40{,}251 < n^A$
without any reallocation.

At wave $t=4$, both designs produce comparable level estimates.
Level coverage is 82.0\%--100\% for the DMM and 90.0\%--100\% for
the classical design across all domain-variable cells. The maximum absolute relative error (MARE) (amongst the strata) ratios (B/A) range from 0.844 to 1.263
with median $\approx 0.99$, confirming comparable point accuracy.
National level coverage is 99\%--100\% for both designs across all
three variables. Level CBI widths are essentially identical
for both designs (Confidence Interval Width (CIW) ratio range 0.993--1.024) --- no material
difference in level precision, as expected given similar sample sizes.

Figure 1 is about here.

\begin{figure}[htbp]
\centering
\centering\input{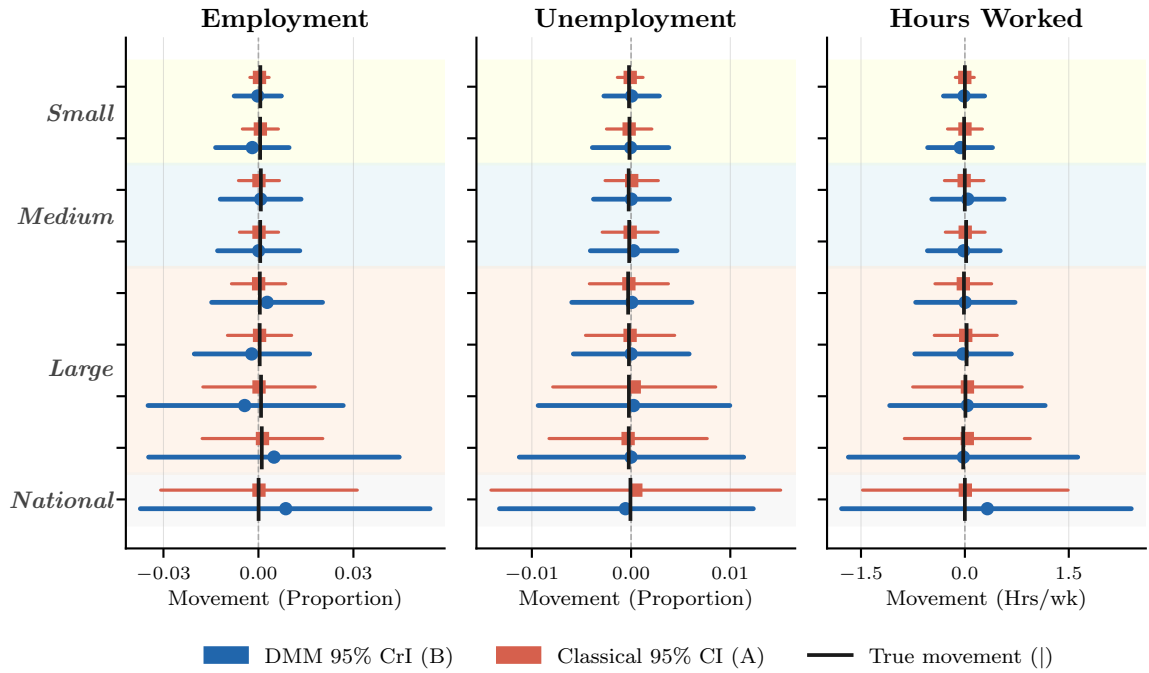}
\caption{Mean movement estimates and 95\% CBI/confidence intervals
at wave $t=4$ across all nine domains for Employment, Unemployment
and Hours Worked ($M=200$ replications). Blue circles (\textbullet): DMM/SHBU Calibrated Bayes interval.
Red squares ($\blacksquare$): classical Patterson-Cochran CI.
Black bar ($|$): true movement. Interval widths differ between the two designs (see text).}
\label{fig:move}
\end{figure}

\paragraph{2.\ Movement estimates (Figure~\ref{fig:move}).}
Figure~\ref{fig:move} shows the most important distinction between
the two frameworks. The DMM movement CBI is 0.88--2.44 times the
width of the classical Patterson-Cochran CI (median ratio $\approx
1.75$) because the CBI includes $\hat{V}_{\mathrm{mod}}$; the
classical CI accounts for sampling variance only.

The movement SE constraints in Table~\ref{tab:constraints} are
design inputs that determine the minimum sample size and overlap
required by the mini-max optimisation; the DMM meets them by
construction. The coverage result in Figure~\ref{fig:coverage} is
a separate inferential property: it measures how well the published
interval covers the true realised movement, which in the simulation
is governed by the AR(1) dynamics.  The DMM CBI achieves 100\%
movement coverage across all 27 domain-variable cells, because it
correctly accounts for $\hat{V}_{\mathrm{mod}}$ --- the model variance
component arising from uncertainty in the AR(1) state at the previous
wave.

The classical CI achieves only 82.0\%--96.0\% movement coverage across
domain-variable cells and 87.5\%--95.0\% nationally.  The classical
design-based CI is theoretically valid for a fixed population at a
single point in time.  When the population itself evolves stochastically
across survey occasions, however, the relevant coverage concept shifts
to one that averages over both sampling variability and population
variability.  The classical CI was designed for the former only; when
evaluated in the latter setting --- as naturally arises in repeated
surveys with evolving populations --- it operates outside its original
scope, and its unconditional coverage over repeated survey occasions
falls below the nominal level accordingly.  This shortfall is
structural: it would persist for any choice of movement magnitude or
$\delta^*$, because it reflects a fundamental difference in inferential
scope between the two frameworks, not a property of the particular
movements simulated here.

For absolute bias at wave $t=4$, the DMM has lower movement bias than the
classical estimator in 9 of 27 domain-variable cells, predominantly
for Unemployment and Hours Worked in national and large domains.
For Employment, classical direct estimation has lower movement bias
in all 9 cells, reflecting the well-known bias-variance trade-off
under HB shrinkage: the posterior mean reduces variance at the cost
of some bias toward the prior mean.

Figure 2 is about here.

\begin{remark}
The movement SE targets $g_{\Delta,d}^{(v)}$ in Table~\ref{tab:constraints}
serve as design inputs for the DMM mini-max optimisation; they are
chosen to be achievable given the calibrated $\hat{V}_{\mathrm{mod},d,t}^{(v)}$
and the planned sample size.  The classical design has its own
movement precision concept: the minimum detectable movement
$\delta^\#_d = Z_{\mathrm{sum}}\sqrt{\hat{V}_{\mathrm{samp},d}^A}$,
which depends only on sampling variance and is well within the
classical framework's capability at $n^A = 42{,}018$.
The two movement precision concepts --- $\delta^*$ for DMM and
$\delta^\#$ for classical --- are not directly comparable because
they target different inferential quantities; $\delta^*$ includes
$\hat{V}_{\mathrm{mod}}$ while $\delta^\#$ does not.
\end{remark}

\begin{figure}[htbp]
\centering
\centering\input{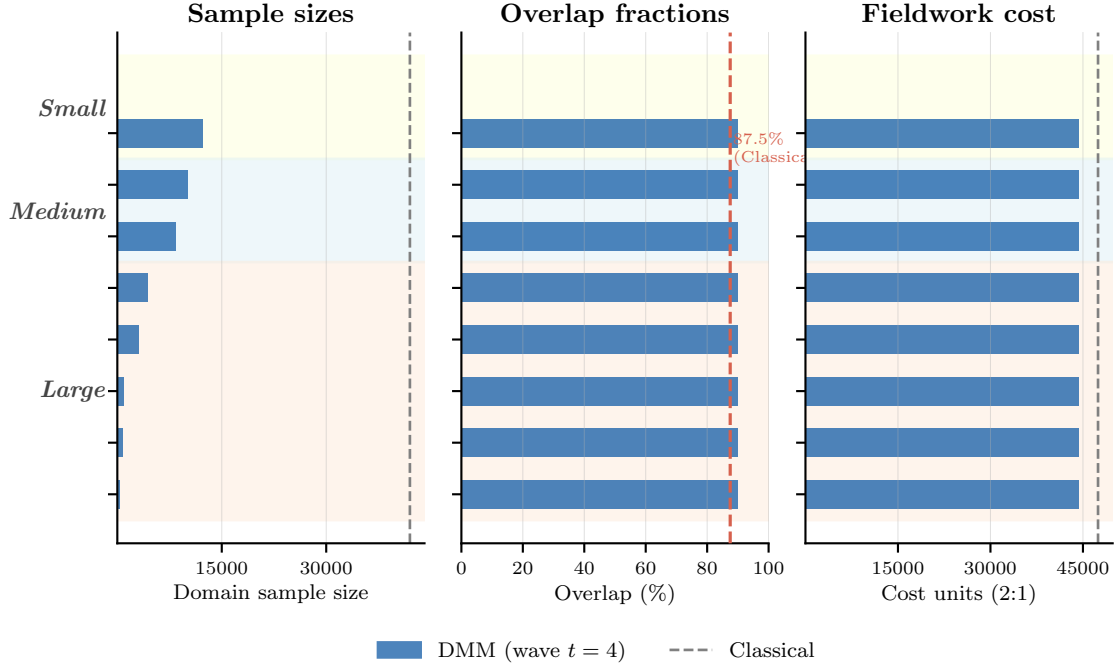}
\caption{Domain sample sizes, overlap fractions and fieldwork costs at
wave $t=4$ for the Dynamic Mini-Max Design (Scenario~B, blue bars).
Dashed lines: classical 7/8 rotation benchmark (sample size and overlap panels only).
Cost panel shows DMM domain-level costs $\text{Cost}_B = 2(n_h^*-m_h^*)+m_h^*$;
no per-domain classical cost benchmark is available since the 6.3\% aggregate saving
compares total DMM cost (44,300) against total classical cost (47,271).
Cost ratio: $c^{\mathrm{new}}:c^{\mathrm{ret}} = 2:1$.
Results at wave $t=3$ are similar (see text).}
\label{fig:design}
\end{figure}

\begin{figure}[htbp]
\centering
\centering\input{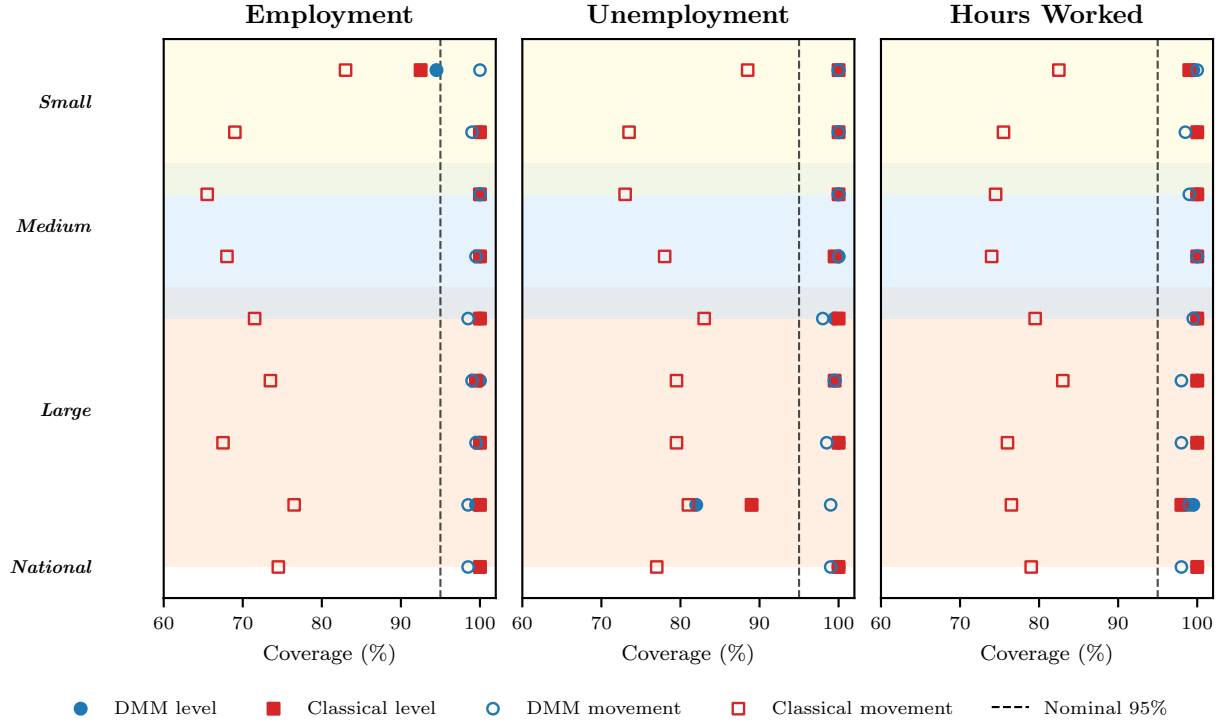}
\caption{Actual versus nominal (95\%) coverage at wave $t=4$ ($M=200$ replications) for Employment, Unemployment and Hours Worked.
Filled blue circles (\textbullet): DMM level coverage. Filled red squares ($\blacksquare$): classical level coverage.
Open blue circles: DMM movement coverage. Open red squares: classical movement coverage.
Vertical dashed line: nominal 95\% coverage.
Background shading: National (white), Large (light orange) = NSW/VIC/QLD/WA, Medium (light blue) = SA/TAS, Small (light yellow) = ACT/NT.}
\label{fig:coverage}
\end{figure}

\paragraph{3.\ Design and cost (Figure~\ref{fig:design}).}
The DMM satisfies all precision constraints at both waves. At wave $t=4$: $n^* = 40{,}251$ versus $n^A = 42{,}018$,
a sample reduction of 1,767 units (4.2\%); overlap
fraction $= 89.9\%$ ($\rho^{\max} = 9/10$ binding); fieldwork cost
$= 44{,}300$ cost units versus the classical $47{,}271$ --- a saving
of 6.3\%. The budget ceiling $B^{\max} = 45{,}000$ does not bind.
The sample reduction arises because the HB linking
and propagation models contribute precision beyond what the sample
alone provides, allowing the optimiser to reduce $n^*$ below $n^A$
while still meeting all constraints.  The joint optimisation simultaneously
pushes $m_h$ to the $\rho^{\max} = 9/10$ respondent burden cap,
further reducing cost by retaining more units at
$c^{\mathrm{ret}} = 1$ rather than recruiting new units at
$c^{\mathrm{new}} = 2$.
Results at wave $t=3$ are structurally identical ($n^*=40{,}184$,
overlap $=89.9\%$, cost $=44{,}227$, saving $=6.4\%$);
Figure~\ref{fig:design} reports wave $t=4$ only.

Figure 3 is about here.

\paragraph{4.\ Actual versus nominal coverage (Figure~\ref{fig:coverage}).}
The figure reveals an asymmetric pattern. DMM level and DMM movement
symbols cluster to the right of the 95\% line --- most DMM coverage
values lie between 97\% and 100\%, well above nominal, with two
exceptions: NSW Unemployment level coverage (82.0\%, reflecting the
marginal level CV failure at 5\% proportional allocation discussed
earlier) and NT Employment level coverage (95.5\%, still above the
nominal 95\%).  Classical level coverage is mostly comparable, though
NSW Unemployment (90.0\%) falls below nominal, also reflecting that
same marginal CV failure.
By contrast, all 27 classical movement symbols (open red squares) lie
to the left of the 95\% line, with coverage ranging from 82.0\%
to 96.0\% across domain-variable cells and 87.5\%--95.0\% nationally.
Over-coverage of the DMM level CBI is consistent with the
tendency for Calibrated Bayes intervals to over-cover when model
parameters are estimated rather than known.

Figure 4 is about here.

\subsubsection{Analysis}
\label{sec:emp_analysis}

The empirical results establish the following findings.

\medskip
\noindent\textbf{1.  More with less.}
Both designs start from the same 5\% proportional allocation of
$n^A = 42{,}018$ units.  The DMM reduces this to $n^* = 40{,}251$
at wave $t=4$ --- a reduction of 1,767 units (4.2\%) --- while meeting
all CV and SE precision targets simultaneously, at a fieldwork cost
of 44,300 versus the classical 47,271 cost units --- a saving of 6.3\%.
The sample reduction is made possible by the HB linking and propagation
models contributing precision beyond what the sample alone provides,
allowing the optimiser to simultaneously reduce $n^*$ and increase
the overlap fraction (89.9\% vs classical 87.5\%) within the
$\rho^{\max} = 9/10$ respondent burden cap.  The movement SE
constraints impose a floor on $n^*$ that is not present in purely
cross-sectional settings; the 4.2\% reduction reflects the
additional binding constraint from $\delta^*$ targets alongside the
level CV targets.

\medskip
\noindent\textbf{2. Level coverage is comparable; movement coverage diverges.}
Both designs produce comparable level point estimates (MARE ratio
0.844--1.263, median $\approx 0.99$) and CBI/confidence interval
widths (CIW ratio 0.993--1.024).  Level coverage is broadly similar
between the two designs, as expected given that both start from
similar sample sizes and the 5\% proportional allocation largely
satisfies the level CV constraints analytically.

Movement coverage, however, diverges sharply.  The DMM achieves
100\% across all 27 domain-variable cells; the classical
design achieves only 82.0\%--96.0\%, with national coverage of
87.5\%--95.0\% across the three variables.
The classical design-based CI is theoretically valid for a fixed
population at a single point in time.  In the repeated-survey setting,
where the true population parameters themselves evolve stochastically
across waves, the relevant coverage concept is unconditional --- averaging
over both sampling variability and population variability.  The
classical CI, designed for the former only, operates outside its
original scope in this setting, and its unconditional coverage falls
below the nominal level accordingly.  This is not a failure of the
classical framework on its own terms, but a consequence of applying
it in a setting that exceeds its inferential scope.

\medskip
\noindent\textbf{3. Additional qualitative advantages.}
Beyond cost and movement coverage, the DMM provides four advantages
the classical framework cannot replicate at any sample size:
\begin{enumerate}
\item \textbf{Optimal estimators.} The SHBU posterior mean minimises
posterior expected squared error for levels and movements
simultaneously; the classical direct
estimator has no equivalent optimality property \citep{godambe1955}.
\item \textbf{Coherent joint posterior: consistent level--movement uncertainty
and valid joint probability statements.}
Since each SHBU draw $b$ produces $\theta_{d,t}^{(v),(b)}$ conditional
on $\theta_{d,t-1}^{(v),(b)}$, the movement
$\Delta_{d,t}^{(v),(b)} = \theta_{d,t}^{(v),(b)} - \theta_{d,t-1}^{(v),(b)}$
is derived from the same draw, so the level and movement credible
intervals share a common posterior and are mutually consistent.  This coherence also enables valid joint probability statements.  For example, the posterior probability that unemployment in a domain
exceeds a policy threshold and is rising is directly computed
as the fraction of the $B$ paired draws
$\bigl(\theta_{d,t}^{(v),(b)},\,\Delta_{d,t}^{(v),(b)}\bigr)$
satisfying both conditions simultaneously; the classical framework,
producing two separate confidence intervals from independent variance
estimates, has no equivalent.
\item \textbf{Sequential updating without intractable or truncated chaining.} Classical composite estimators \citep[Chapter~12]{cochran1977} chain
estimates across waves by regressing on the immediately preceding wave
only, because extending the chain to all previous waves $t=1,\ldots,T$
requires estimating and maintaining an increasing number of regression
coefficients and becomes operationally intractable as $T$ grows.
\citet{cochran1977} notes this results in some loss of precision,
which is accepted in practice.
The SHBU avoids this truncation: the AR(1) prior encodes the full
posterior history in $p(\theta_{t-1} \mid\text{y}_{1:t-1})$,
which propagates forward automatically at each wave as the prior for
$\theta_t$; no manual chaining decisions or truncation are required
regardless of how large $T$ becomes.

\item \textbf{Small area estimation.} The HB linking model borrows
strength across strata, enabling reliable estimates for sparse
domains where the direct estimator would require a substantially
larger sample to achieve the same precision. This allows the DMM
to meet precision targets for small domains at a sample size
$n^* < n^A$ that would be insufficient under a purely
design-based approach.

\end{enumerate}

\medskip
\noindent\textbf{4.  Sensitivity to AR(1) misspecification.}
When $\hat{\phi}^{(v)} = 0$ (i.e. model is misspecified), level coverage degrades
to 88\%--96\% nationally at waves $t=3,4$, while movement
coverage remains high (99\%--100\%) because the cross-sectional
branch is used. Correct AR(1) calibration is therefore a prerequisite
for valid level inference. Full sensitivity results are in the
Supplementary Materials.

\begin{remark}
The DMM framework assumes a stationary AR(1) propagation model with
a correctly calibrated coefficient $\hat{\phi}^{(v)}$. The sensitivity
analysis above shows that misspecification causes substantial coverage
degradation. This is a limitation of the current framework: if the
true process is non-stationary, exhibits structural breaks, or follows
a higher-order autoregression, the AR(1) assumption will be violated
and inference may be unreliable. Robustifying the propagation model
against such misspecification is an important direction for future research.  Note that when transisting from the classical to the DMM framework, the access to $K$ historical
waves of classical direct estimates enable the NSO not only to calibrate the AR(1)
parameters via the moment estimator, but also to
select the best-fitting temporal model using standard time series methods. The AR(1) assumption
is therefore not blindly imposed but empirically testable from the
historical data which can mitigate
against misspecification of the propogation model.
\end{remark}
\section{Concluding Remarks}
\label{sec:conclusion}

This paper extends the Mini-Max HB framework of \citet{tam2026a} to
repeated surveys through two methodological advances: a DMM framework comprising a Dynamic Mini-Max
Design jointly determining $\{n_{h,t}\}$ and $\{m_{h,t}\}$ at each wave
to minimise cost subject to simultaneous CV and SE precision constraints for all target variables and domains, and budget and respondent load restrictions; and an
SHBU propagating posterior uncertainty sequentially across waves.

Both designs begin from the same 5\% proportional allocation of
$n^A = 42{,}018$ units.  Analytically, this allocation largely satisfies
the level CV precision constraints; the only marginal exceptions are
NT and ACT Unemployment, where the achieved CV of approximately 27\%
marginally exceeds the 25\% target --- a gap that a classical Bethel
reallocation would correct with a trivial sample increase.

The simulation shows several findings of interest.
First, the DMM reduces the sample to $n^* = 40{,}251$ at wave $t=4$
--- a 4.2\% reduction from $n^A$ --- while meeting all precision,
budget and respondent load targets, at a fieldwork cost of 44,300
versus the classical 47,271 cost units, giving a cost saving of 6.3\%.
The modest sample reduction, relative to purely cross-sectional
applications of the Mini-Max framework, reflects the additional
binding constraint from movement SE targets alongside level CV targets;
the static and dynamic frameworks address different constraint sets and direct comparison
of achieved reductions is not meaningful.

Second, level accuracy and coverage are comparable
between designs (MARE ratio 0.844--1.263, level CIW ratio 0.993--1.024),
as expected given similar sample sizes and the analytical result that
both designs largely satisfy the level CV constraints.

Third, movement coverage diverges sharply: the DMM achieves
100\% across all 27 domain-variable cells while the
classical design achieves only 82.0\%--96.0\% (87.5\%--95.0\%
nationally).  The classical design-based CI is valid for a fixed
population; in a repeated-survey setting where population parameters
evolve stochastically, its unconditional coverage falls below the
nominal level because it was designed for a different inferential scope.
The DMM CBI, by explicitly incorporating $\hat{V}_{\mathrm{mod}}$,
correctly targets the unconditional coverage and achieves it.

Fourth, the DMM provides additional advantages the classical
framework cannot replicate at any sample size: optimal Bayesian
estimators; coherent joint uncertainty for levels and movements
enabling valid joint probability statements; and automatic sequential
updating without intractable composite estimator chaining.

\medskip

The framework assumes a stationary AR(1) propagation model with a
correctly calibrated coefficient $\hat{\phi}^{(v)}$. The sensitivity
analysis shows that misspecifying $\hat{\phi}^{(v)} = 0$ can severely
degrade level coverage for large domains: at wave $t=4$, NSW
Unemployment drops to 37.5\%, VIC Employment to 56.5\%, and NSW
Hours Worked to 65.0\%, compared with 82.0\%--100.0\% under the
correctly calibrated SHBU.  In practice, the availability of
$K$ historical waves of classical direct estimates enables the NSO
not only to calibrate the AR(1) parameters via the moment estimator,
but also to select the best-fitting temporal model using standard
time series methods, mitigating against misspecification.
Extensions include robustifying the propagation model specification
and hyperparameter estimation and extending the movement estimator from wave-to-wave change
to quarterly and annual aggregates, which require accumulating
posterior draws across multiple waves.
\clearpage

\clearpage

\section*{Appendix 1: Proof of the Moment Estimator for $\sigma_\eta^{2(v)}$}
\label{app:moment}

Under the stationary AR(1) propagation \eqref{eq:prop_continuous}, the true
domain state satisfies:
\begin{equation}
    \theta_{d,t}^{(v)} - \theta_{d,t-1}^{(v)}
    = (\phi^{(v)} - 1)\left(\theta_{d,t-1}^{(v)} - \mu^{(v)}\right)
    + \eta_{d,t}^{(v)},
    \label{eq:app_diff}
\end{equation}
where $\eta_{d,t}^{(v)} \sim \mathcal{N}(0, \sigma_\eta^{2(v)})$. Taking the
variance of \eqref{eq:app_diff} and using independence of $\eta_{d,t}^{(v)}$
and $\theta_{d,t-1}^{(v)}$:
\begin{equation}
    \mathrm{Var}\!\left(\theta_{d,t}^{(v)} - \theta_{d,t-1}^{(v)}\right)
    = (1 - \phi^{(v)})^2\, \mathrm{Var}\!\left(\theta_{d,t-1}^{(v)}\right)
    + \sigma_\eta^{2(v)}.
    \label{eq:app_var1}
\end{equation}
Under stationarity ($|\phi^{(v)}| < 1$), the marginal variance of the state is:
\begin{equation}
    \mathrm{Var}\!\left(\theta_{d,t}^{(v)}\right)
    = \frac{\sigma_\eta^{2(v)}}{1 - \phi^{2(v)}}
    = \frac{\sigma_\eta^{2(v)}}{(1-\phi^{(v)})(1+\phi^{(v)})}.
    \label{eq:app_statvar}
\end{equation}
Substituting \eqref{eq:app_statvar} into \eqref{eq:app_var1}:
\begin{align}
    \mathrm{Var}\!\left(\theta_{d,t}^{(v)} - \theta_{d,t-1}^{(v)}\right)
    &= \frac{(1-\phi^{(v)})^2\, \sigma_\eta^{2(v)}}{(1-\phi^{(v)})(1+\phi^{(v)})}
    + \sigma_\eta^{2(v)} \notag \\
    &= \frac{(1-\phi^{(v)})\,\sigma_\eta^{2(v)}}{1+\phi^{(v)}}
    + \sigma_\eta^{2(v)} \notag \\
    &= \sigma_\eta^{2(v)}\left[\frac{1-\phi^{(v)}}{1+\phi^{(v)}} + 1\right]
    = \frac{2\,\sigma_\eta^{2(v)}}{1+\phi^{(v)}}.
    \label{eq:app_diffvar}
\end{align}

The direct survey estimate satisfies $\hat{\theta}_{d,t}^{(v)} = \theta_{d,t}^{(v)}
+ e_{d,t}^{(v)}$, where $e_{d,t}^{(v)}$ is the sampling error with variance
$\psi_{d,t}^{(v)}$. When the survey uses overlapping samples, the sampling errors
at adjacent waves are correlated through the matched units. By the
Patterson--Cochran covariance formula \citep{patterson1950, cochran1977}:
\begin{equation}
    \mathrm{Cov}\!\left(e_{d,t}^{(v)},\, e_{d,t-1}^{(v)}\right)
    = \frac{m_{h,t}}{n_{h,t}\, n_{h,t-1}}\,\mathrm{DEFF}_h^{(v)}\,
    S_h^{2(v)}\,\rho_h^{(v)},
    \label{eq:app_pcov}
\end{equation}
where $\rho_h^{(v)}$ is the between-wave correlation of the characteristic for
matched units. The variance of the observed first
difference is therefore:
\begin{equation}
    \mathrm{Var}\!\left(\hat{\theta}_{d,t}^{(v)} - \hat{\theta}_{d,t-1}^{(v)}\right)
    = \frac{2\,\sigma_\eta^{2(v)}}{1+\phi^{(v)}}
    + \psi_{d,t}^{(v)} + \psi_{d,t-1}^{(v)}
    - 2\,\frac{m_{h,t}}{n_{h,t}\, n_{h,t-1}}\,\mathrm{DEFF}_h^{(v)}\,
    S_h^{2(v)}\,\rho_h^{(v)}.
    \label{eq:app_obsvar}
\end{equation}

Equation~\eqref{eq:app_obsvar} holds for each adjacent wave pair
$(k-1, k)$, $k = 2,\ldots,K$. Averaging both sides over the $K-1$
pairs, and noting that $\sigma_\eta^{2(v)}$ and $\phi^{(v)}$ are
constant across waves:
\begin{equation}
    \frac{1}{K-1}\sum_{k=2}^{K}
    \mathrm{Var}\!\left(\hat{\theta}_{d,k}^{(v)} -
    \hat{\theta}_{d,k-1}^{(v)}\right)
    = \frac{2\,\sigma_\eta^{2(v)}}{1+\phi^{(v)}}
    + \bar{\psi}_d^{(v),\mathrm{corr}},
    \label{eq:app_avgvar}
\end{equation}
where $\bar{\psi}_d^{(v),\mathrm{corr}}$ is the average of the sampling
variance and covariance terms across wave pairs, defined in
\eqref{eq:psi_corr}.

From $K$ historical waves, equate the sample variance of the observed first
differences to its theoretical counterpart \eqref{eq:app_obsvar}, averaging the
sampling variance and covariance terms over adjacent wave pairs:
\begin{equation}
    \frac{1}{K-1}\sum_{k=2}^{K}\left(\hat{\theta}_{d,k}^{(v)}
    - \hat{\theta}_{d,k-1}^{(v)}\right)^2
    = \frac{2\,\sigma_\eta^{2(v)}}{1+\phi^{(v)}} + \bar{\psi}_d^{(v),\mathrm{corr}},
    \label{eq:app_mom}
\end{equation}
where $\bar{\psi}_d^{(v),\mathrm{corr}}$ is the corrected average sampling
variance defined in equation~\eqref{eq:psi_corr}. Solving \eqref{eq:app_mom} for
$\sigma_\eta^{2(v)}$ and replacing $\phi^{(v)}$ with its pilot estimate
$\hat{\phi}^{(v)}$ gives the general moment estimator \eqref{eq:moment_scale}.

Note that setting $\hat{\rho}_h^{(v)} = 0$ in \eqref{eq:psi_corr} gives
$\bar{\psi}_d^{(v),\mathrm{corr}} = \bar{\psi}_d^{(v)}$, the unadjusted average
sampling variance. This corresponds to using only the unmatched (fresh)
units at each wave to compute $\hat{\theta}_{d,k}^{(v)}$. The moment
estimator then simplifies to:
\begin{equation}
    \hat{s}_\eta^{2(v)}\big|_{\rho=0} = \frac{1+\hat{\phi}^{(v)}}{2}
    \left[\frac{1}{K-1}\sum_{k=2}^{K}
    \left(\hat{\theta}_{d,k}^{(v),\mathrm{new}} - \hat{\theta}_{d,k-1}^{(v),\mathrm{new}}\right)^2
    - \bar{\psi}_d^{(v),\mathrm{new}}\right],
    \label{eq:app_rho0}
\end{equation}
where $\hat{\theta}_{d,k}^{(v),\mathrm{new}}$ is computed from the fresh units
only and $\bar{\psi}_d^{(v),\mathrm{new}}$ uses the correspondingly larger
sampling variances $\mathrm{DEFF}_h^{(v)} S_h^{2(v)} / (n_{h,k} - m_{h,k})$.
\section*{Appendix 2: Proof of the Model Variance Estimator}
\label{app:modvar}

Under the AR(1) propagation model \eqref{eq:prop_continuous}, conditional on the
wave $t-1$ state $\theta_{d,t-1}^{(v)}$ and the AR(1) parameters
$\bm{\xi}^{(v)} = (\phi^{(v)}, \sigma_\eta^{2(v)}, \mu^{(v)})$:
\begin{equation}
    \theta_{d,t}^{(v)} \mid \theta_{d,t-1}^{(v)}, \bm{\xi}^{(v)} \;\sim\;
    \mathcal{N}\!\left(\mu^{(v)} + \phi^{(v)}(\theta_{d,t-1}^{(v)} - \mu^{(v)}),\;
    \sigma_\eta^{2(v)}\right).
    \label{eq:app2_cond}
\end{equation}
Therefore the movement $\Delta_{d,t}^{(v)} = \theta_{d,t}^{(v)} -
\theta_{d,t-1}^{(v)}$ has conditional distribution:
\begin{equation}
    \Delta_{d,t}^{(v)} \mid \theta_{d,t-1}^{(v)}, \bm{\xi}^{(v)} \;\sim\;
    \mathcal{N}\!\left((\phi^{(v)}-1)(\theta_{d,t-1}^{(v)} - \mu^{(v)}),\;
    \sigma_\eta^{2(v)}\right).
    \label{eq:app2_delta}
\end{equation}
The conditional mean and variance are:
\begin{align}
    E\!\left[\Delta_{d,t}^{(v)} \mid \theta_{d,t-1}^{(v)}, \bm{\xi}^{(v)}\right]
    &= (\phi^{(v)}-1)(\theta_{d,t-1}^{(v)} - \mu^{(v)})
    = \hat{\Delta}_{d,t}^{(v)}, \label{eq:app2_mean}\\
    \mathrm{Var}\!\left(\Delta_{d,t}^{(v)} \mid \theta_{d,t-1}^{(v)},
    \bm{\xi}^{(v)}\right) &= \sigma_\eta^{2(v)}. \label{eq:app2_var}
\end{align}

The predictive variance of $\Delta_{d,t}^{(v)}$ given the wave $t-1$ posterior
$\pi(\theta_{d,t-1}^{(v)}, \bm{\xi}^{(v)} \mid y_{1:t-1})$ is:
\begin{align}
    V_{\mathrm{mod},d,t}^{(v)}
    &= \mathrm{Var}\!\left(\Delta_{d,t}^{(v)} \mid y_{1:t-1}\right) \notag\\
    &= \underbrace{E\!\left[\mathrm{Var}\!\left(\Delta_{d,t}^{(v)} \mid
    \theta_{d,t-1}^{(v)}, \bm{\xi}^{(v)}\right)\right]}_{\text{Term 1}}
    + \underbrace{\mathrm{Var}\!\left(E\!\left[\Delta_{d,t}^{(v)} \mid
    \theta_{d,t-1}^{(v)}, \bm{\xi}^{(v)}\right]\right)}_{\text{Term 2}},
    \label{eq:app2_totvar}
\end{align}
where both expectations and the variance are taken over the wave $t-1$ posterior.

\noindent Term 1 — expectation of the conditional variance
\eqref{eq:app2_var}:
\begin{equation}
    E\!\left[\sigma_\eta^{2(v)}\right] \approx \frac{1}{B}\sum_{b=1}^{B}
    \sigma_\eta^{2(v),(b)},
    \label{eq:app2_term1}
\end{equation}
estimated by averaging the draw-specific innovation variances from the wave $t-1$
AR(1) parameter draws from the SHBU.

\noindent Term 2 --- variance of the conditional mean \eqref{eq:app2_mean}.
Each draw $b$ gives a draw-specific conditional mean
$\tilde{\Delta}_{d,t}^{(v),(b)} = (\phi^{(v),(b)}-1)(\hat{\theta}_{d,t-1}^{(v),(b)} -
\mu^{(v),(b)})$ --- a deterministic plug-in prediction, not a draw from
any distribution. The sample variance of these across $B$ draws is:
\begin{equation}
    \mathrm{Var}\!\left(\hat{\Delta}_{d,t}^{(v)}\right) \approx
    \frac{1}{B-1}\sum_{b=1}^{B}\left[\tilde{\Delta}_{d,t}^{(v),(b)} -
    \tilde{\Delta}_{d,t}^{(v),\mathrm{HB}}\right]^2,
    \label{eq:app2_term2}
\end{equation}
where $\tilde{\Delta}_{d,t}^{(v),\mathrm{HB}} = \frac{1}{B}\sum_{b=1}^B
\tilde{\Delta}_{d,t}^{(v),(b)}$.

Substituting \eqref{eq:app2_term1} and \eqref{eq:app2_term2} into
\eqref{eq:app2_totvar} gives the estimator \eqref{eq:vmod}:
\begin{equation*}
    \hat{V}_{\mathrm{mod},d,t}^{(v)} = \frac{1}{B}\sum_{b=1}^{B}
    \sigma_\eta^{2(v),(b)} + \frac{1}{B-1}\sum_{b=1}^{B}
    \left[\tilde{\Delta}_{d,t}^{(v),(b)} -
    \tilde{\Delta}_{d,t}^{(v),\mathrm{HB}}\right]^2. \qed
\end{equation*}

\medskip Note that Term~1 is the average innovation variance across
draws --- the irreducible movement noise that cannot be reduced by the design.
Term~2 is the variance of the draw-specific one-step-ahead movement predictions
--- the additional uncertainty arising from not knowing the true state
$\theta_{d,t-1}^{(v)}$ and the AR(1) parameters $(\phi^{(v)}, \mu^{(v)})$
exactly. Both terms are directly computable from the wave $t-1$ AR(1) parameter draws
$\{(\phi^{(v),(b)}, \sigma_\eta^{2(v),(b)}, \mu^{(v),(b)},
\hat{\theta}_{d,t-1}^{(v),(b)})\}_{b=1}^B$.

\end{document}


\maketitle
\tableofcontents
\newpage

These supplementary tables support the four figures in the main paper.
All results are at wave $t=4$, $M=200$ replications, $B=2{,}000$ draws.
DMM settings: $\rho^{\max} = 9/10$, $B^{\max} = 45{,}000$ cost units.

\section{Tables for Figure 1: Level Estimates}
\subsection{Employment}
\begin{tabular}{lrrrrrrr}
\toprule
Domain & Truth & Est$_B$ & CI$^L_B$ & CI$^U_B$ & Est$_A$ & CI$^L_A$ & CI$^U_A$ \\
\midrule
National & 0.6083 & 0.6069 & 0.5982 & 0.6155 & 0.6069 & 0.5984 & 0.6154 \\
NSW & 0.5785 & 0.5811 & 0.5674 & 0.5947 & 0.5814 & 0.5681 & 0.5948 \\
VIC & 0.6272 & 0.6232 & 0.6083 & 0.6382 & 0.6228 & 0.6081 & 0.6374 \\
QLD & 0.6122 & 0.6074 & 0.5919 & 0.6229 & 0.6074 & 0.5920 & 0.6228 \\
WA & 0.6394 & 0.6363 & 0.6153 & 0.6572 & 0.6366 & 0.6157 & 0.6575 \\
SA & 0.5984 & 0.5984 & 0.5763 & 0.6204 & 0.5981 & 0.5764 & 0.6199 \\
TAS & 0.5808 & 0.5734 & 0.5352 & 0.6116 & 0.5724 & 0.5345 & 0.6103 \\
ACT & 0.6937 & 0.6908 & 0.6437 & 0.7380 & 0.6907 & 0.6438 & 0.7377 \\
NT & 0.6379 & 0.6727 & 0.6161 & 0.7293 & 0.6749 & 0.6189 & 0.7309 \\
\bottomrule
\end{tabular}

\medskip
\noindent Est$_B$: DMM posterior mean. CI$^{L/U}_B$: DMM 95\% CBI bounds.
Est$_A$: classical direct estimate. CI$^{L/U}_A$: classical 95\% CI bounds.

\subsection{Unemployment}
\begin{tabular}{lrrrrrrr}
\toprule
Domain & Truth & Est$_B$ & CI$^L_B$ & CI$^U_B$ & Est$_A$ & CI$^L_A$ & CI$^U_A$ \\
\midrule
National & 0.0343 & 0.0351 & 0.0319 & 0.0384 & 0.0348 & 0.0316 & 0.0380 \\
NSW & 0.0322 & 0.0354 & 0.0310 & 0.0399 & 0.0351 & 0.0306 & 0.0395 \\
VIC & 0.0345 & 0.0357 & 0.0312 & 0.0401 & 0.0353 & 0.0310 & 0.0397 \\
QLD & 0.0361 & 0.0347 & 0.0295 & 0.0398 & 0.0344 & 0.0293 & 0.0394 \\
WA & 0.0366 & 0.0358 & 0.0287 & 0.0429 & 0.0356 & 0.0285 & 0.0426 \\
SA & 0.0343 & 0.0334 & 0.0266 & 0.0402 & 0.0329 & 0.0261 & 0.0396 \\
TAS & 0.0380 & 0.0337 & 0.0202 & 0.0471 & 0.0333 & 0.0198 & 0.0467 \\
ACT & 0.0268 & 0.0305 & 0.0171 & 0.0439 & 0.0298 & 0.0166 & 0.0430 \\
NT & 0.0388 & 0.0370 & 0.0223 & 0.0517 & 0.0370 & 0.0223 & 0.0517 \\
\bottomrule
\end{tabular}

\medskip
\noindent Est$_B$: DMM posterior mean. CI$^{L/U}_B$: DMM 95\% CBI bounds.
Est$_A$: classical direct estimate. CI$^{L/U}_A$: classical 95\% CI bounds.

\subsection{Hours Worked}
\begin{tabular}{lrrrrrrr}
\toprule
Domain & Truth & Est$_B$ & CI$^L_B$ & CI$^U_B$ & Est$_A$ & CI$^L_A$ & CI$^U_A$ \\
\midrule
National & 21.1724 & 21.1887 & 20.8365 & 21.5408 & 21.1744 & 20.8300 & 21.5189 \\
NSW & 20.1297 & 20.3184 & 19.7443 & 20.8925 & 20.3141 & 19.7452 & 20.8831 \\
VIC & 21.6432 & 21.5786 & 20.9444 & 22.2128 & 21.5480 & 20.9268 & 22.1692 \\
QLD & 21.5650 & 21.4327 & 20.7644 & 22.1010 & 21.4213 & 20.7534 & 22.0893 \\
WA & 22.5646 & 22.5075 & 21.5889 & 23.4260 & 22.5054 & 21.5861 & 23.4247 \\
SA & 20.2483 & 20.1660 & 19.2240 & 21.1081 & 20.1447 & 19.2043 & 21.0850 \\
TAS & 19.3615 & 19.4182 & 17.8238 & 21.0125 & 19.3943 & 17.7916 & 20.9969 \\
ACT & 24.8811 & 24.7254 & 22.6143 & 26.8365 & 24.6910 & 22.5846 & 26.7973 \\
NT & 23.9182 & 25.5348 & 22.6918 & 28.3778 & 25.5695 & 22.7071 & 28.4319 \\
\bottomrule
\end{tabular}

\medskip
\noindent Est$_B$: DMM posterior mean. CI$^{L/U}_B$: DMM 95\% CBI bounds.
Est$_A$: classical direct estimate. CI$^{L/U}_A$: classical 95\% CI bounds.

\section{Tables for Figure 2: Movement Estimates}
\subsection{Employment}
\begin{tabular}{lrrrrrrr}
\toprule
Domain & Truth & Est$_B$ & CI$^L_B$ & CI$^U_B$ & Est$_A$ & CI$^L_A$ & CI$^U_A$ \\
\midrule
National & 0.00059 & -0.00022 & -0.00767 & 0.00727 & 0.00033 & -0.00271 & 0.00337 \\
NSW & 0.00054 & -0.00194 & -0.01354 & 0.00969 & 0.00061 & -0.00508 & 0.00630 \\
VIC & 0.00076 & 0.00071 & -0.01207 & 0.01347 & 0.00016 & -0.00631 & 0.00663 \\
QLD & 0.00056 & 0.00007 & -0.01292 & 0.01308 & 0.00018 & -0.00600 & 0.00636 \\
WA & 0.00045 & 0.00276 & -0.01478 & 0.02026 & 0.00006 & -0.00852 & 0.00863 \\
SA & 0.00040 & -0.00212 & -0.02026 & 0.01620 & 0.00036 & -0.00979 & 0.01050 \\
TAS & 0.00082 & -0.00434 & -0.03487 & 0.02680 & 0.00019 & -0.01756 & 0.01793 \\
ACT & 0.00106 & 0.00493 & -0.03470 & 0.04448 & 0.00128 & -0.01778 & 0.02034 \\
NT & 0.00005 & 0.00864 & -0.03730 & 0.05422 & 0.00017 & -0.03088 & 0.03122 \\
\bottomrule
\end{tabular}

\subsection{Unemployment}
\begin{tabular}{lrrrrrrr}
\toprule
Domain & Truth & Est$_B$ & CI$^L_B$ & CI$^U_B$ & Est$_A$ & CI$^L_A$ & CI$^U_A$ \\
\midrule
National & -0.00021 & 0.00005 & -0.00274 & 0.00286 & -0.00009 & -0.00138 & 0.00121 \\
NSW & -0.00016 & -0.00005 & -0.00390 & 0.00380 & -0.00019 & -0.00248 & 0.00210 \\
VIC & -0.00024 & 0.00003 & -0.00379 & 0.00388 & 0.00006 & -0.00262 & 0.00275 \\
QLD & -0.00019 & 0.00026 & -0.00411 & 0.00463 & -0.00009 & -0.00291 & 0.00273 \\
WA & -0.00030 & 0.00008 & -0.00598 & 0.00614 & -0.00022 & -0.00421 & 0.00376 \\
SA & -0.00021 & 0.00001 & -0.00582 & 0.00585 & -0.00010 & -0.00460 & 0.00439 \\
TAS & -0.00022 & 0.00026 & -0.00937 & 0.00996 & 0.00032 & -0.00790 & 0.00853 \\
ACT & -0.00023 & -0.00000 & -0.01127 & 0.01135 & -0.00030 & -0.00827 & 0.00768 \\
NT & -0.00006 & -0.00057 & -0.01327 & 0.01232 & 0.00048 & -0.01413 & 0.01508 \\
\bottomrule
\end{tabular}

\subsection{Hours Worked}
\begin{tabular}{lrrrrrrr}
\toprule
Domain & Truth & Est$_B$ & CI$^L_B$ & CI$^U_B$ & Est$_A$ & CI$^L_A$ & CI$^U_A$ \\
\midrule
National & -0.00001 & -0.01174 & -0.30937 & 0.28550 & -0.00125 & -0.13804 & 0.13555 \\
NSW & -0.01110 & -0.06630 & -0.53796 & 0.39926 & 0.00122 & -0.25072 & 0.25315 \\
VIC & -0.00132 & 0.04118 & -0.47784 & 0.56693 & -0.01038 & -0.29553 & 0.27476 \\
QLD & 0.01827 & -0.01602 & -0.54112 & 0.51085 & 0.00488 & -0.28254 & 0.29230 \\
WA & -0.01353 & 0.00462 & -0.71074 & 0.72489 & -0.02252 & -0.43192 & 0.38688 \\
SA & 0.02502 & -0.02758 & -0.72378 & 0.67209 & 0.01220 & -0.44235 & 0.46676 \\
TAS & 0.00764 & 0.03513 & -1.08582 & 1.15919 & 0.03684 & -0.75349 & 0.82717 \\
ACT & -0.02387 & -0.02115 & -1.68226 & 1.62922 & 0.03491 & -0.87405 & 0.94387 \\
NT & -0.00038 & 0.32465 & -1.78089 & 2.40250 & 0.00650 & -1.47735 & 1.49036 \\
\bottomrule
\end{tabular}

\section{Tables for Figure 3: Design and Cost}
\subsection{Wave $t=3$}
\begin{tabular}{lrrrr}
\toprule
Domain & $n_h^*$ & $m_h^*$ & Overlap\% & Cost$_B$ \\
\midrule
NSW & 12,324 & 11,085 & 89.9 & 13,563 \\
VIC & 9,930 & 8,931 & 89.9 & 10,929 \\
QLD & 8,418 & 7,570 & 89.9 & 9,266 \\
SA & 3,088 & 2,778 & 90.0 & 3,398 \\
WA & 4,350 & 3,912 & 89.9 & 4,788 \\
TAS & 943 & 848 & 89.9 & 1,038 \\
NT & 358 & 322 & 89.9 & 394 \\
ACT & 773 & 695 & 89.9 & 851 \\
\midrule
\textbf{Total (DMM)} & \textbf{40,184} & \textbf{36,141} & \textbf{89.9} & \textbf{44,227} \\
\textbf{Classical ($n^A=42{,}018$)} & & & \textbf{87.5} & \textbf{47,271} \\
\textbf{Saving (\%)} & & & & \textbf{6.4} \\
\bottomrule
\end{tabular}

\medskip
\noindent $n_h^*$: DMM domain sample size. $m_h^*$: retained units.
Overlap\% $= m_h^*/n_h^* \times 100$.
Cost$_B = 2(n_h^*-m_h^*)+m_h^*$.
Classical benchmark: $n^A=42{,}018$ with 7/8 rotation.
Binding constraint: $\rho^{\max}=9/10$; budget ceiling $B^{\max}=45{,}000$ does not bind.

\subsection{Wave $t=4$}
\begin{tabular}{lrrrr}
\toprule
Domain & $n_h^*$ & $m_h^*$ & Overlap\% & Cost$_B$ \\
\midrule
NSW & 12,271 & 11,038 & 90.0 & 13,504 \\
VIC & 10,090 & 9,074 & 89.9 & 11,106 \\
QLD & 8,362 & 7,521 & 89.9 & 9,203 \\
SA & 3,094 & 2,783 & 89.9 & 3,405 \\
WA & 4,358 & 3,919 & 89.9 & 4,797 \\
TAS & 944 & 849 & 89.9 & 1,039 \\
NT & 358 & 322 & 89.9 & 394 \\
ACT & 774 & 696 & 89.9 & 852 \\
\midrule
\textbf{Total (DMM)} & \textbf{40,251} & \textbf{36,202} & \textbf{89.9} & \textbf{44,300} \\
\textbf{Classical ($n^A=42{,}018$)} & & & \textbf{87.5} & \textbf{47,271} \\
\textbf{Saving (\%)} & & & & \textbf{6.3} \\
\bottomrule
\end{tabular}

\medskip
\noindent $n_h^*$: DMM domain sample size. $m_h^*$: retained units.
Overlap\% $= m_h^*/n_h^* \times 100$.
Cost$_B = 2(n_h^*-m_h^*)+m_h^*$.
Classical benchmark: $n^A=42{,}018$ with 7/8 rotation.
Binding constraint: $\rho^{\max}=9/10$; budget ceiling $B^{\max}=45{,}000$ does not bind.

\section{Tables for Figure 4: Actual versus Nominal Coverage}
\subsection{Employment}
\begin{tabular}{lrrrr}
\toprule
Domain & Level Cov$_B$ & Level Cov$_A$ & Move Cov$_B$ & Move Cov$_A$ \\
\midrule
National & 100.0 & 100.0 & 100.0 & 87.5 \\
NSW & 100.0 & 99.5 & 100.0 & 89.0 \\
VIC & 99.5 & 99.5 & 100.0 & 90.5 \\
QLD & 99.0 & 100.0 & 100.0 & 89.0 \\
WA & 100.0 & 100.0 & 100.0 & 82.5 \\
SA & 100.0 & 100.0 & 100.0 & 82.0 \\
TAS & 100.0 & 100.0 & 100.0 & 84.0 \\
ACT & 100.0 & 100.0 & 100.0 & 86.0 \\
NT & 95.5 & 92.5 & 100.0 & 88.0 \\
\bottomrule
\end{tabular}

\medskip
\noindent Nominal coverage = 95\%. Cov$_B$: DMM CBI. Cov$_A$: classical CI.

\subsection{Unemployment}
\begin{tabular}{lrrrr}
\toprule
Domain & Level Cov$_B$ & Level Cov$_A$ & Move Cov$_B$ & Move Cov$_A$ \\
\midrule
National & 99.0 & 100.0 & 100.0 & 95.0 \\
NSW & 82.0 & 90.0 & 100.0 & 95.0 \\
VIC & 100.0 & 100.0 & 100.0 & 93.0 \\
QLD & 99.5 & 99.0 & 100.0 & 96.0 \\
WA & 100.0 & 100.0 & 100.0 & 92.0 \\
SA & 100.0 & 100.0 & 100.0 & 95.0 \\
TAS & 100.0 & 100.0 & 100.0 & 86.5 \\
ACT & 100.0 & 100.0 & 100.0 & 91.0 \\
NT & 100.0 & 100.0 & 100.0 & 95.0 \\
\bottomrule
\end{tabular}

\medskip
\noindent Nominal coverage = 95\%. Cov$_B$: DMM CBI. Cov$_A$: classical CI.

\subsection{Hours Worked}
\begin{tabular}{lrrrr}
\toprule
Domain & Level Cov$_B$ & Level Cov$_A$ & Move Cov$_B$ & Move Cov$_A$ \\
\midrule
National & 100.0 & 99.5 & 100.0 & 89.5 \\
NSW & 100.0 & 98.0 & 100.0 & 92.0 \\
VIC & 100.0 & 100.0 & 100.0 & 94.0 \\
QLD & 100.0 & 100.0 & 100.0 & 90.5 \\
WA & 100.0 & 100.0 & 100.0 & 84.0 \\
SA & 100.0 & 100.0 & 100.0 & 89.5 \\
TAS & 100.0 & 100.0 & 100.0 & 89.5 \\
ACT & 100.0 & 100.0 & 100.0 & 92.0 \\
NT & 99.0 & 99.0 & 100.0 & 88.5 \\
\bottomrule
\end{tabular}

\medskip
\noindent Nominal coverage = 95\%. Cov$_B$: DMM CBI. Cov$_A$: classical CI.

\section{Sensitivity to AR(1) Misspecification ($\hat{\phi}^{(v)}=0$)}

Each sub-table reports level and movement coverage under (i) calibrated
$\hat{\phi}^{(v)}$ (SHBU) and (ii) misspecified $\hat{\phi}^{(v)}=0$
($\phi=0$). $M=200$ replications, $B=2{,}000$ draws.

\subsection{Employment}
\begin{tabular}{llrrrrrr}
\toprule
Wave & Domain & \multicolumn{2}{c}{Level Cov (\%)} & \multicolumn{2}{c}{Movement Cov (\%)} & \multicolumn{2}{c}{MARE} \\
\cmidrule(lr){3-4}\cmidrule(lr){5-6}\cmidrule(lr){7-8}
& & SHBU & $\phi{=}0$ & SHBU & $\phi{=}0$ & SHBU & $\phi{=}0$ \\
\midrule
2 & National & 99.5 & 96.0 & 100.0 & 99.5 & 0.33 & 0.33 \\
 & NSW & 97.5 & 77.0 & 99.5 & 100.0 & 0.72 & 1.04 \\
 & VIC & 99.0 & 55.5 & 97.5 & 73.5 & 0.65 & 1.40 \\
 & QLD & 100.0 & 98.0 & 100.0 & 100.0 & 0.67 & 0.60 \\
 & WA & 100.0 & 99.5 & 100.0 & 100.0 & 0.60 & 0.69 \\
 & SA & 100.0 & 100.0 & 100.0 & 100.0 & 0.74 & 0.81 \\
 & TAS & 100.0 & 100.0 & 100.0 & 90.0 & 2.56 & 1.43 \\
 & ACT & 100.0 & 98.5 & 98.5 & 99.0 & 1.30 & 1.46 \\
 & NT & 99.0 & 77.0 & 100.0 & 100.0 & 3.93 & 4.80 \\
\midrule
3 & National & 100.0 & 95.5 & 100.0 & 100.0 & 0.33 & 0.33 \\
 & NSW & 99.5 & 80.0 & 100.0 & 100.0 & 0.93 & 1.10 \\
 & VIC & 99.0 & 62.0 & 100.0 & 100.0 & 0.76 & 1.40 \\
 & QLD & 100.0 & 98.0 & 100.0 & 100.0 & 0.76 & 0.61 \\
 & WA & 100.0 & 97.5 & 100.0 & 100.0 & 0.97 & 0.74 \\
 & SA & 100.0 & 99.5 & 99.5 & 100.0 & 0.74 & 0.84 \\
 & TAS & 100.0 & 99.5 & 97.5 & 100.0 & 0.99 & 1.57 \\
 & ACT & 100.0 & 99.0 & 100.0 & 100.0 & 1.58 & 1.56 \\
 & NT & 98.5 & 83.0 & 100.0 & 100.0 & 4.16 & 4.56 \\
\midrule
4 & National & 100.0 & 95.0 & 100.0 & 100.0 & 0.38 & 0.36 \\
 & NSW & 100.0 & 84.5 & 100.0 & 100.0 & 0.63 & 1.02 \\
 & VIC & 99.5 & 56.5 & 100.0 & 100.0 & 0.75 & 1.49 \\
 & QLD & 99.0 & 93.0 & 100.0 & 100.0 & 0.85 & 0.70 \\
 & WA & 100.0 & 99.0 & 100.0 & 100.0 & 0.72 & 0.72 \\
 & SA & 100.0 & 98.5 & 100.0 & 100.0 & 0.60 & 0.83 \\
 & TAS & 100.0 & 97.5 & 100.0 & 100.0 & 1.52 & 1.51 \\
 & ACT & 100.0 & 98.5 & 100.0 & 100.0 & 1.37 & 1.54 \\
 & NT & 95.5 & 80.0 & 100.0 & 100.0 & 5.45 & 4.61 \\
\midrule
\bottomrule
\end{tabular}

\subsection{Unemployment}
\begin{tabular}{llrrrrrr}
\toprule
Wave & Domain & \multicolumn{2}{c}{Level Cov (\%)} & \multicolumn{2}{c}{Movement Cov (\%)} & \multicolumn{2}{c}{MARE} \\
\cmidrule(lr){3-4}\cmidrule(lr){5-6}\cmidrule(lr){7-8}
& & SHBU & $\phi{=}0$ & SHBU & $\phi{=}0$ & SHBU & $\phi{=}0$ \\
\midrule
2 & National & 99.5 & 96.0 & 98.5 & 99.0 & 2.16 & 2.17 \\
 & NSW & 94.5 & 51.0 & 99.5 & 100.0 & 5.70 & 8.03 \\
 & VIC & 99.5 & 95.5 & 99.5 & 99.0 & 2.45 & 3.14 \\
 & QLD & 97.0 & 79.0 & 99.0 & 99.0 & 4.48 & 5.75 \\
 & WA & 100.0 & 82.0 & 98.5 & 100.0 & 3.93 & 6.45 \\
 & SA & 99.5 & 93.0 & 99.5 & 100.0 & 4.34 & 4.61 \\
 & TAS & 99.0 & 73.5 & 99.5 & 96.0 & 10.19 & 14.23 \\
 & ACT & 99.0 & 92.0 & 99.0 & 98.5 & 15.60 & 14.28 \\
 & NT & 100.0 & 100.0 & 100.0 & 100.0 & 7.06 & 6.67 \\
\midrule
3 & National & 100.0 & 91.5 & 100.0 & 100.0 & 2.69 & 2.68 \\
 & NSW & 76.5 & 41.0 & 98.5 & 100.0 & 9.64 & 9.00 \\
 & VIC & 100.0 & 90.0 & 100.0 & 100.0 & 3.38 & 3.94 \\
 & QLD & 97.5 & 86.0 & 100.0 & 100.0 & 5.35 & 5.03 \\
 & WA & 100.0 & 86.5 & 100.0 & 100.0 & 4.60 & 5.78 \\
 & SA & 99.5 & 97.5 & 100.0 & 100.0 & 4.36 & 3.93 \\
 & TAS & 100.0 & 71.0 & 86.0 & 100.0 & 12.82 & 13.01 \\
 & ACT & 100.0 & 91.0 & 100.0 & 100.0 & 13.90 & 16.84 \\
 & NT & 100.0 & 99.5 & 100.0 & 100.0 & 6.32 & 6.41 \\
\midrule
4 & National & 99.0 & 88.0 & 100.0 & 100.0 & 3.17 & 3.17 \\
 & NSW & 82.0 & 37.5 & 100.0 & 100.0 & 10.05 & 9.57 \\
 & VIC & 100.0 & 87.0 & 100.0 & 100.0 & 3.96 & 4.65 \\
 & QLD & 99.5 & 87.5 & 100.0 & 100.0 & 4.55 & 4.58 \\
 & WA & 100.0 & 90.0 & 100.0 & 100.0 & 4.70 & 5.34 \\
 & SA & 100.0 & 98.0 & 100.0 & 100.0 & 4.37 & 3.94 \\
 & TAS & 100.0 & 78.5 & 100.0 & 100.0 & 11.86 & 12.38 \\
 & ACT & 100.0 & 90.0 & 100.0 & 100.0 & 15.20 & 17.80 \\
 & NT & 100.0 & 100.0 & 100.0 & 100.0 & 6.65 & 6.08 \\
\midrule
\bottomrule
\end{tabular}

\subsection{Hours Worked}
\begin{tabular}{llrrrrrr}
\toprule
Wave & Domain & \multicolumn{2}{c}{Level Cov (\%)} & \multicolumn{2}{c}{Movement Cov (\%)} & \multicolumn{2}{c}{MARE} \\
\cmidrule(lr){3-4}\cmidrule(lr){5-6}\cmidrule(lr){7-8}
& & SHBU & $\phi{=}0$ & SHBU & $\phi{=}0$ & SHBU & $\phi{=}0$ \\
\midrule
2 & National & 99.5 & 97.0 & 100.0 & 100.0 & 0.38 & 0.38 \\
 & NSW & 99.0 & 70.5 & 98.0 & 100.0 & 0.79 & 1.53 \\
 & VIC & 100.0 & 58.5 & 100.0 & 90.0 & 0.71 & 1.79 \\
 & QLD & 100.0 & 96.5 & 100.0 & 100.0 & 0.68 & 0.69 \\
 & WA & 100.0 & 97.5 & 100.0 & 99.5 & 0.79 & 0.82 \\
 & SA & 100.0 & 99.5 & 100.0 & 100.0 & 0.85 & 0.88 \\
 & TAS & 100.0 & 99.5 & 100.0 & 99.5 & 2.34 & 1.44 \\
 & ACT & 99.5 & 98.0 & 97.0 & 97.5 & 1.75 & 1.80 \\
 & NT & 100.0 & 71.5 & 100.0 & 100.0 & 2.87 & 5.68 \\
\midrule
3 & National & 100.0 & 97.0 & 99.0 & 99.5 & 0.44 & 0.43 \\
 & NSW & 97.0 & 63.5 & 100.0 & 100.0 & 1.27 & 1.71 \\
 & VIC & 100.0 & 67.5 & 99.5 & 100.0 & 0.76 & 1.72 \\
 & QLD & 100.0 & 97.0 & 100.0 & 100.0 & 0.75 & 0.69 \\
 & WA & 100.0 & 99.0 & 100.0 & 100.0 & 0.85 & 0.89 \\
 & SA & 100.0 & 98.5 & 100.0 & 100.0 & 0.82 & 0.96 \\
 & TAS & 100.0 & 98.5 & 98.5 & 100.0 & 1.27 & 1.60 \\
 & ACT & 100.0 & 96.0 & 99.5 & 100.0 & 1.84 & 1.87 \\
 & NT & 100.0 & 74.5 & 99.0 & 100.0 & 5.43 & 5.37 \\
\midrule
4 & National & 100.0 & 95.5 & 100.0 & 100.0 & 0.37 & 0.37 \\
 & NSW & 100.0 & 65.0 & 100.0 & 100.0 & 1.01 & 1.66 \\
 & VIC & 100.0 & 69.0 & 100.0 & 100.0 & 0.65 & 1.73 \\
 & QLD & 100.0 & 95.0 & 100.0 & 99.5 & 0.78 & 0.65 \\
 & WA & 100.0 & 97.0 & 100.0 & 100.0 & 0.78 & 0.87 \\
 & SA & 100.0 & 98.0 & 100.0 & 100.0 & 0.86 & 0.91 \\
 & TAS & 100.0 & 99.5 & 100.0 & 100.0 & 1.22 & 1.52 \\
 & ACT & 100.0 & 97.0 & 100.0 & 100.0 & 1.86 & 1.88 \\
 & NT & 99.0 & 73.0 & 100.0 & 100.0 & 6.76 & 5.48 \\
\midrule
\bottomrule
\end{tabular}

\section{Table for MARE and CIW Ratios: Level Accuracy at Wave $t=4$}

This table supports the claim in the main paper that the MARE ratios
(DMM/classical) range from 0.844 to 1.263 with median $\approx 0.99$,
and that the CIW ratios range from 0.993 to 1.024.
All results are at wave $t=4$, $M=200$ replications, $B=2{,}000$ draws.

\begin{table}[ht]
\centering
\caption{Level Accuracy at Wave $t=4$: MARE and CIW Ratios (DMM / Classical).
         MARE = maximum absolute relative error amongst strata (\%).
         CIW = 95\% interval width.
         Ratio $<1$ favours DMM; ratio $>1$ favours classical.}
\label{tab:mare_ciw}
\small
\begin{tabular}{llrrrrr}
\toprule
Domain & Variable
  & MARE$_B$ & MARE$_A$ & MARE ratio
  & CIW$_B$ & CIW ratio \\
\midrule
\multicolumn{7}{l}{\textit{National}} \\
National & Employment      & 0.38 & 0.37 & 1.027 & 0.0173 & 1.024 \\
National & Hours Worked    & 0.37 & 0.40 & 0.925 & 0.7044 & 1.022 \\
National & Unemployment    & 3.17 & 2.51 & \textbf{1.263} & 0.0065 & 1.016 \\
\addlinespace
\multicolumn{7}{l}{\textit{Large domains}} \\
NSW      & Employment      & 0.63 & 0.66 & 0.955 & 0.0272 & 1.019 \\
NSW      & Hours Worked    & 1.01 & 0.99 & 1.020 & 1.1482 & 1.009 \\
NSW      & Unemployment    & 10.05 & 8.95 & 1.123 & 0.0089 & 1.000 \\
VIC      & Employment      & 0.75 & 0.80 & 0.938 & 0.0299 & 1.024 \\
VIC      & Hours Worked    & 0.65 & 0.72 & 0.903 & 1.2684 & 1.021 \\
VIC      & Unemployment    & 3.96 & 3.31 & 1.196 & 0.0089 & 1.011 \\
QLD      & Employment      & 0.85 & 0.84 & 1.012 & 0.0310 & 1.006 \\
QLD      & Hours Worked    & 0.78 & 0.83 & 0.940 & 1.3366 & 1.000 \\
QLD      & Unemployment    & 4.55 & 5.16 & 0.882 & 0.0102 & 1.000 \\
WA       & Employment      & 0.72 & 0.71 & 1.014 & 0.0420 & 1.005 \\
WA       & Hours Worked    & 0.78 & 0.85 & 0.918 & 1.8371 & 0.999 \\
WA       & Unemployment    & 4.70 & 4.82 & 0.975 & 0.0142 & 1.007 \\
\addlinespace
\multicolumn{7}{l}{\textit{Medium domains}} \\
SA       & Employment      & 0.60 & 0.57 & 1.053 & 0.0442 & 1.016 \\
SA       & Hours Worked    & 0.86 & 0.83 & 1.036 & 1.8841 & 1.002 \\
SA       & Unemployment    & 4.37 & 5.18 & \textbf{0.844} & 0.0136 & 1.015 \\
TAS      & Employment      & 1.52 & 1.61 & 0.944 & 0.0764 & 1.008 \\
TAS      & Hours Worked    & 1.22 & 1.28 & 0.953 & 3.1887 & 0.995 \\
TAS      & Unemployment    & 11.86 & 12.90 & 0.919 & 0.0269 & 1.000 \\
\addlinespace
\multicolumn{7}{l}{\textit{Small domains}} \\
ACT      & Employment      & 1.37 & 1.32 & 1.038 & 0.0944 & 1.005 \\
ACT      & Hours Worked    & 1.86 & 1.72 & 1.081 & 4.2222 & 1.002 \\
ACT      & Unemployment    & 15.20 & 12.95 & 1.174 & 0.0268 & 1.015 \\
NT       & Employment      & 5.45 & 5.80 & 0.940 & 0.1131 & 1.010 \\
NT       & Hours Worked    & 6.76 & 6.90 & 0.980 & 5.6860 & 0.993 \\
NT       & Unemployment    & 6.65 & 6.42 & 1.036 & 0.0293 & 0.997 \\
\midrule
\multicolumn{3}{l}{Range}
  & & 0.844--1.263 & & 0.993--1.024 \\
\multicolumn{3}{l}{Median}
  & & \multicolumn{1}{r}{0.980} & & \multicolumn{1}{r}{1.009} \\
\bottomrule
\end{tabular}

\medskip
\noindent\small
MARE$_B$: DMM posterior mean MARE (\%). MARE$_A$: classical direct estimator MARE (\%).
MARE ratio $=$ MARE$_B$/MARE$_A$.
CIW$_B$: DMM 95\% CBI width. CIW ratio $=$ CIW$_B$/CIW$_A$.
Bold entries mark the minimum (0.844, SA Unemployment) and maximum (1.263,
National Unemployment) MARE ratios.
Classical benchmark: $n^A = 42{,}018$ with 7/8 rotation.
\end{table}